\documentclass[aps,pra,twocolumn, groupedaddress,showpacs]{revtex4}
\usepackage{amsmath}
\usepackage{hyperref}
\pdfoutput=1

	\usepackage[pdftex]{graphicx}

\newcommand*{\req}[1]{(\ref{#1})}
\newcommand*{\un}[1]{\ensuremath{\ \mathrm{#1}}}

\newcommand*{\uvec}[1]{\ensuremath{\mathbf{\hat{#1}}}}
\renewcommand*{\vec}[1]{\ensuremath{\mathbf{#1}}}

\newcommand*{\ket}[1]{\ensuremath{\left|{#1}\right\rangle}}

\newcommand*{\etal}[0]{{\em et al}}

	\DeclareGraphicsExtensions{.pdf, .jpg, .tif}

\begin{document}

\title{Scalability of quantum computation with addressable optical lattices }


\author{T. R. Beals}
\affiliation{Department of Physics, University of California, Berkeley, California 94720, USA}

\author{J. Vala}
\altaffiliation{Current address: Department of of Mathematical Physics, National University of Ireland, Maynooth, Co. Kildare, Ireland}
\affiliation{Department of Chemistry and Pitzer Center for Theoretical Chemistry, University of California, Berkeley, California 94720, USA}

\author{K. B. Whaley}
\email{whaley@berkeley.edu}
\affiliation{Department of Chemistry and Pitzer Center for Theoretical Chemistry, University of California, Berkeley, California 94720, USA}

\date{\today}

\begin{abstract}
We make a detailed analysis of error mechanisms, gate fidelity, and scalability of proposals for quantum computation with neutral atoms in addressable (large lattice constant) optical lattices. 
We have identified possible limits to the size of quantum computations, arising in 3D optical lattices from current limitations on the ability to perform single qubit gates in parallel and in 2D lattices from constraints on laser power.
Our results suggest that 3D arrays as large as  $100 \times 100\times 100$ sites (i.e., $\sim 10^6$ qubits) may be achievable, provided two-qubit gates can be performed with sufficiently high precision and degree of parallelizability.  Parallelizability of long range interaction-based two-qubit gates is qualitatively compared to that of collisional gates. Different methods of performing single qubit gates are compared, and a lower bound of $1 \times 10^{-5}$ is determined on the error rate for the error mechanisms affecting $^{133}$Cs in a blue-detuned lattice with Raman transition-based single qubit gates, given reasonable limits on experimental parameters.  
\end{abstract}

\pacs{03.67.Lx, 02.70.Hm, 37.10.Jk, 32.80.Qk}

\maketitle

\section{Introduction}\label{sec:intro}
Neutral atoms trapped in optical lattices constitute a promising system for quantum information processing~\cite{deutsch:control,deutsch:neutral}. Single qubit operations and qubit readout have already been demonstrated~\cite{schrader:150501}, albeit in a non-scalable system, and a number of two-qubit gates have been proposed~\cite{jaksch:2208, jaksch:1975}. Addressable optical lattices---in which the lattice spacing is large enough that individual lattice sites can be targeted by a laser~\cite{scheunemann:051801}---offer an environment that can be scaled to thousands of qubits in a three-dimensional (3D) array~\cite{weiss:040302}. Preparation, loading, and imaging of an addressable optical lattice have recently been demonstrated~\cite{nelson:lattice}.

Achieving large-scale fault-tolerant quantum computation requires single and two-qubit gates with extremely high fidelity, as well as the ability to perform many gates in parallel. Current estimates of the fault tolerance error threshold range from  $10^{-3}$ to $10^{-7}$ for conventional quantum error correction, depending on the difficulty of communication between physically distant qubits and the ability to prepare certain states ``offline'' in a reliable manner~\cite{szkopek:0411111,cheng:012320,aliferis:0504218,svore:0604090,steane:042322}. More radical error correction schemes~\cite{raussendorf:0510135,knill:042322,knill:2005vh,reichardt:1170522} may offer better thresholds, but at the cost of high overhead. For the optical lattice system that is the focus of this work, the most relevant estimates~\cite{svore:0604090} suggest a fault tolerance threshold on the order of $10^{-5}$.

While much research has focused on schemes for realization of qubits and performing quantum operations in a wide variety of experimental systems, the detailed physics and scalability of specific architectures have to date received less attention.  Some work has been done on system-level analysis of architectural issues~\cite{oskin:976922,kubiatowicz:spaa,copsey:1263787,isailovic:interconnection}.  However, to provide real numbers to questions such as how large a system of qubits may be made and how many quantum operations can be made on this, it is necessary to undertake detailed analysis of the behavior of the proposed qubits in situ.   The different physics involved in different implementations pose a variety of challenges, some of which may be system-specific while others, such as achieving gate fidelities with fault tolerant threshold values are quite generic.  For example, the need to use thermal ensembles rather than pure states---as is usual in gas state and solid state proposals---has presented a major challenge for liquid state NMR because of limitations imposed by small thermal polarization and difficulties of initialization~\cite{warren:09121997,braunstein:9811018,jones:nmr,boykin:alg_cool}.

For qubits defined in internal states of trapped neutral atoms or trapped ions, both the qubit interactions and their environmental decoherence mechanisms are well understood~\cite{deslauriers:103007}.  This enables architectural issues to be examined with full microscopic analysis of all physical features of such qubits.   For ion traps, analysis of the limiting features of trapped ion physics led to the proposal of a modular, multiplexed trap architecture to allow scale up from a few ( $< 10$) ions to many, possibly thousands of individually addressable ions~\cite{wineland:ions,kielpinski:scaleup2002,schaetz:scalableions} and progress demonstrating components of such an architecture for small numbers of ions is now underway~\cite{rowe:257,barrett:nature2004,hensinger:Ttrap,stick:surface,seidelin:surfacetrap2006}.  For neutral atoms  trapped in optical lattices, up to 250 individual atomic qubits have been trapped and imaged in an optical lattice system that can be readily scaled up to include thousands of atoms and whose spacing is large enough to allow individual addressability~\cite{nelson:lattice}.  Large arrays of sublattice addressable trapped atoms have also been made~\cite{lee:dw,anderlini:sublattice}.  Trapped neutral atoms and ions are somewhat complementary; individual addressing and quantum gates are more straightforward to implement with ions and have already been demonstrated experimentally for small numbers of ions (e.g., forming entangled states of $ \leq 8$ ions~\cite{leibfried:cat,haffner:ions} and implementing algorithms with two~\cite{brickman:050306} and three qubits~\cite{chiaverini:602}) but 
achieving
scaleup to even hundreds of ions still presents a serious technical challenge. In contrast, although scaleup to lattices containing hundreds of neutral atoms has been demonstrated in an addressable system~\cite{nelson:lattice} and gates have been demonstrated in non-addressable systems~\cite{anderlini:sublattice} and in addressable dipole traps~\cite{yavuz:0509176,jones:040301}, they have not yet been accomplished in lattices containing individually addressable atoms. Very recently, techniques for addressing individual sites in a lattice with small spacing have been proposed~\cite{gorshkov:093005,cho:020502}.
 
In contrast to the situation for atomic and ionic qubits, progress in scaleup of solid state realizations of qubits---such as Josephson junctions or $^{31}$P in Si---from pairs to many qubits is in a far more rudimentary state.  The kind of detailed analysis that is required to develop specific physical devices cannot be made yet, since the underlying microscopic physics of the qubits when in situ is not currently well enough understood, although significant progress is being made~\cite{clark:silicon}. Furthermore, while solid state systems are often generically referred to as `scalable' because of the ability to fabricate large scale solid state devices, 
the individual elements or qubits  are not as reproducible as gas phase qubits due to the complexities and variability of their surroundings~\cite{keyes:737}.  Nevertheless, architectural studies are beginning to be made for these systems~\cite{hollenberg:set,hollenberg:045311}.

In this work, we go a step further in analysis of scaleup
for trapped neutral atoms,
undertaking a detailed physical investigation of the effects of single qubit errors and other imperfections that limit the scalability of neutral atom quantum computation in an individually addressable optical lattice.   We consider different candidate single qubit gates and their sensitivity to various sources of experimental error.  We then compare this to calculations of the threshold rate for fault tolerant computation under the appropriate conditions, in order to estimate how large a quantum computation may be made within current technological constraints and possible near-term improvements.   To our knowledge, this is the first such detailed physical estimation of a practical limit on physical scaleup for any proposed experimental implementation of pure state quantum computation.  We hope that this detailed analysis for trapped neutral atoms will spur similar analyses for other physical implementations once the relevant microscopic physics is better understood.  Given that no experimental system will have unlimited scalability, such physical analysis of technical limits to scalable systems of functioning qubits within current technology is an essential complement to theoretical algorithmic scaling characteristics  derived from complexity theory.

We restrict our analysis here to addressable optical lattices.  While we do not explicitly consider the alternative lattices with global addressing that are also being studied experimentally~\cite{porto:neutral}, we shall make some comments at the end of this paper on relative benefits that these other lattices might offer.  The analysis in this paper employs a combination of perturbation theory and numerical techniques such as the pseudo-spectral method with a Chebychev decomposition of the Schr\"odinger propagator~\cite{kosloff:propagation} to quantify the effects of both memory and gate errors deriving from all known sources for trapped atoms.
In some respects our calculations complement and extend those of Saffman and Walker~\cite{saffman:022347} for $^{87}$Rb atoms in dipole traps. However, that work did not address the issue of scalability that is 
analyzed here after the various error rates have been quantified.
When a specific choice of parameters is necessary, we consider here $^{133}$Cs atoms in an addressable optical lattice~\cite{nelson:lattice} of lattice constant $a=5\ \mu$m, with depth of $U_L = 200 \un{\mu K}$. The lattice is ortho-rhombic in geometry, and is created by blue-detuned beams at 800 nm, with an intrapair angle of $\sim 9^\circ$ for each of the three pairs of beams. For the one-dimensional case, the lattice potential is given by:
\begin{eqnarray}
V(x) = \frac{U_L}{2} \cos\left(\frac{2 \pi}{a} x \right). \label{eq:latticepot}
\end{eqnarray}
Field-insensitive sub-levels of the $6s\ ^2S_{1/2}$ hyperfine ground-state manifold are chosen as the qubit basis: $\ket{0} \equiv \ket{F=3, m_F=0}$ and $\ket{1} \equiv \ket{F=4, m_F=0}$. The auxiliary levels $\ket{2} \equiv \ket{F=4, m_F=1}$ and $\ket{3} \equiv \ket{F=3, m_F=1}$ are also involved in the single qubit gate presented here (see Fig. \ref{fig:gate}). The procedure for loading and initializing the lattice is described in detail in Weiss \etal ~\cite{weiss:040302}, and Vala \etal ~\cite{vala:0307085}. We assume that atoms are cooled to the vibrational ground state, e.g., with 3D Raman sideband cooling~\cite{kerman:439,han:724}. 

This rest of this paper is organized as follows. Section \ref{sec:lattice} contains an analysis of the error mechanisms due to the lattice itself, such as scattering processes and loss of atoms from the lattice. Section \ref{sec:ramansinglequbit} analyzes the single qubit gate proposal based on a Raman two-photon process and Section \ref{sec:microwavesinglequbit} analyzes microwave pulse-based single qubit gates.  We consider the effects of 
off-resonant transitions of non-target atoms, scattering, heating of target atoms, and addressing beam targeting and intensity errors for both types of single qubit gate.
Section \ref{sec:twoqubit} contains a brief comparative discussion of the scalability of different classes of two-qubit gates. Section \ref{sec:analysis} provides an analysis of the results from the previous sections and their collective implications for the scalability of quantum computation in an addressable optical lattice. In section \ref{sec:conclusion}, we summarize our conclusions, discuss some possible ways to bypass the limitations identified here and identify some useful applications within the constraints established here.

\section{Error Mechanisms\label{sec:error_mech}} 
Fault tolerance thresholds are sometimes expressed in terms of a ``unified'' error rate comprising both storage and gate errors, but are often also written in terms of separate gate and storage error rates (and occasionally also preparation and read-out error rates).  In order for error correction to be able to keep up with storage errors, a practical quantum computer must be able to perform many gates in parallel. Consequently, estimates of error threshold values have typically assumed that gates can be performed on arbitrarily many qubits in parallel~\cite{steane:042322}. Storage errors can then be considered on a 
similar footing to gate errors that occur with a frequency given by multiplying the storage error rate by the 
typical gate time, $T_1$ to obtain an effective `error per gate time' that can be combined with the error per gate (EPG) in analysis of overall error rates.

Fault-tolerance threshold theorems assume maximal parallelizability~\cite{steane:042322}, implying that all or nearly all qubits can be addressed simultaneously ($n_A \simeq N$, where $N$ is the total number of physical qubits and $n_A$ is the number that may be simultaneously addressed). 
In most proposed schemes for quantum computing this is extremely hard, except for the trivial case where one desires to perform the exact same gate on all atoms simultaneously. 
Parallelizability thus constitutes an important figure of merit, since if $n_A$ grows slower than $N$, the effective storage error rate will eventually exceed the capacity of any error correction protocol.  If only a fraction $n_A/N$ of qubits can be addressed, the effective EPG for a storage error will be approximately equal to the storage error rate multiplied by the ratio $N T_1 /n_A$. As we show in Section \ref{sec:analysis}, $n_A$ is on the order of $N^{2/3}$ in 3D lattices, while in 2D lattices $n_A$ can be on the order of $N$. $T_1$ is on the order of tens of microseconds for the microwave pulse-based single qubit gate, whereas with sufficient laser power, it can be nanoseconds or less for the Raman single qubit gate.

In the remainder of this section we derive expressions for the EPG for various decoherence mechanisms encountered by atoms trapped in an addressable optical lattice, using $^{133}$Cs as a specific example where necessary.

\subsection{Optical lattice-induced storage errors\label{sec:lattice}}
We assume that the lattice has already been prepared, and that each lattice site is initially occupied by exactly one atom in the motional $\ket{0}$ state. A detailed description of a procedure to achieve this perfectly loaded lattice is contained in Vala \etal ~\cite{vala:0307085}.

\subsubsection{Photon Scattering\label{sec:latticescatter}}
\begin{figure}[pt] \centering
\includegraphics[width=3.5 in,  keepaspectratio=true]{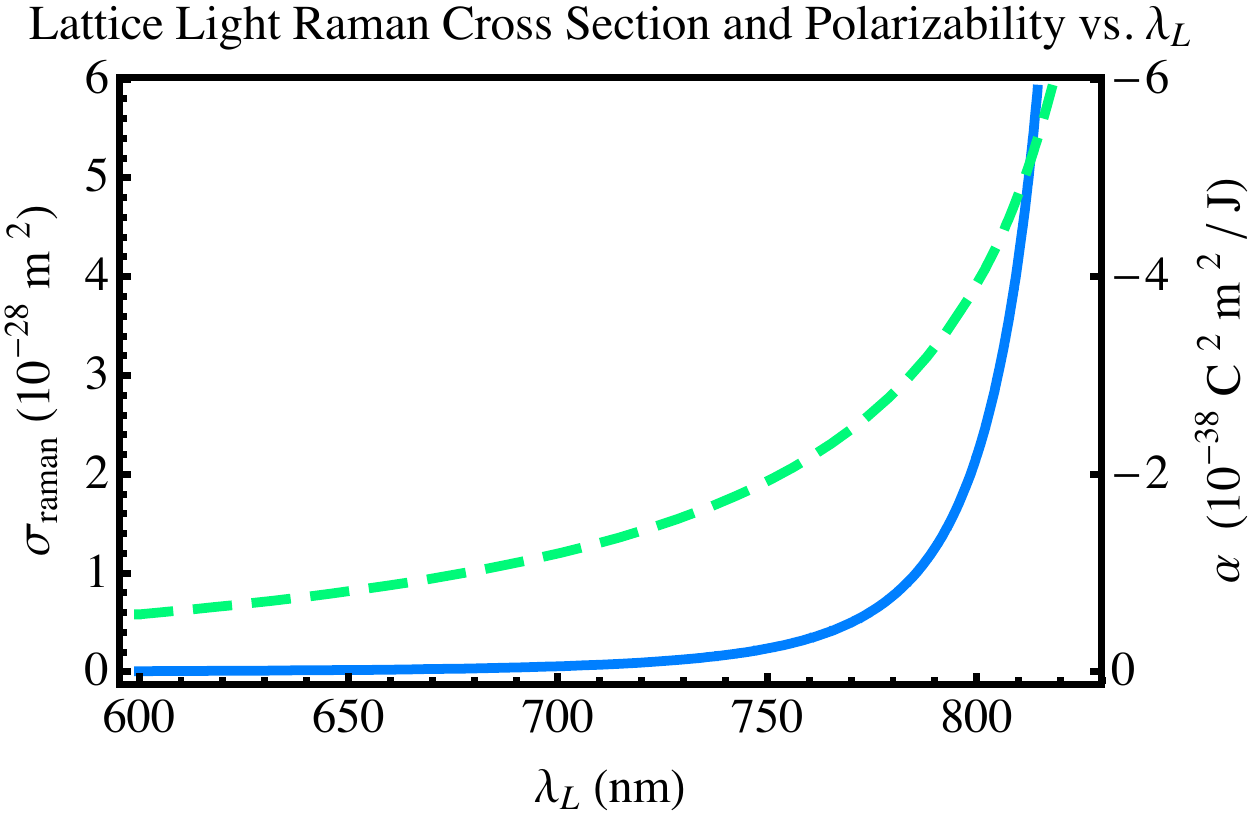}
\caption{\label{fig:scatter} (Color online) The solid blue line shows the Raman scattering cross-section 
$\sigma(\omega_L)$ for a $^{133}$Cs atom in a blue-detuned optical lattice, as a function of lattice light wavelength, $\lambda_L$. The dashed green line shows the frequency-dependent polarizability 
$\alpha(\omega_L)$ at the frequency $\omega_L = 2 \pi c / \lambda_L$. Both $\sigma(\omega_L)$ and $\alpha(\omega_L)$ are calculated for $\epsilon_{+1}$ polarized light interacting with the initial state $\ket{F=3,m_F=0}$, with results for the $\ket{F=4,m_F=0}$ state or opposite polarization light being similar.}
\end{figure}
Both Raman scattering, in which the initial and final states of the atom differ, and Rayleigh scattering, in which they do not, are sources of storage errors. Fortunately, the decohering effects of Rayleigh scattering can be partially suppressed with pulse sequences~\cite{andersen:013405}. For Raman scattering, no such method exists, and so we focus our analysis here on this form of scattering. The effective storage EPG due to Raman scattering is given by $\frac{N}{n_A} \frac{T _1\Gamma}{\hbar}$, where $T_1$ is the gate time and $\Gamma/\hbar$ the scattering rate. 
Calculating the relative transition strengths (see Appendix the details), we find that roughly half of the errors induced by Raman scattering will be bit-flip errors, while the rest will be leakage to non-qubit states. The Raman cross section $\sigma(\omega_L)$ (where $\omega_L$ is the lattice light frequency) can be calculated as described in the Appendix, and is shown in Fig. \ref{fig:scatter} as a function of the lattice light wavelength. The scattering rate is related to the cross section by 
\begin{eqnarray}
\Gamma_{L}/\hbar & = & \frac{c \epsilon_0 \overline{E^2}}{2 \hbar \omega_L} \sigma(\omega_L),  \label{eq:scatterrate}
\end{eqnarray}
where $\overline{E^2}$ is average over an atomic spatial distribution of the peak electric field squared.
We note that the optical lattice potential depth is given by $U_{L}(\omega_L, E_0^2) =  \frac{E_0^2}{4}\left|\alpha(\omega_L)\right|$ \footnote{See Eq. (16) in N. B. Delone and V. P. Krainov, \emph{Physics--Uspekhi} \textbf{42}, 669 (1999).}.
The polarizability is shown as a function of wavelength in Fig. \ref{fig:scatter}. For an atom in the ground state in a red-detuned lattice $\overline{E^2} \simeq E_0^2$, whereas in a blue-detuned lattice, $\overline{E^2} = \frac{\hbar \pi^2}{2 a^2 m \omega_\tau} E_0^2$, where $\omega_\tau =\frac{\pi}{a} \sqrt{2 U_L / m}$ is the characteristic trapping frequency. 
We can then calculate the Raman scattering rate for the blue-detuned and red-detuned cases:
\begin{subequations}
\begin{eqnarray}
\Gamma_{\text{blue}}/\hbar & = & \frac{\pi c \epsilon_0}{a \omega_L}  \sqrt{\frac{U_L(\omega_L,E_0^2)}{2 m}} \frac{\sigma(\omega_L)}{\left| \alpha(\omega_L)\right|} \label{eq:bluescatter} \\
\Gamma_{\text{red}}/\hbar & = & \frac{2 c \epsilon_0}{\hbar \omega_L} U_L(\omega_L,E_0^2)  \frac{\sigma(\omega_L)}{\left| \alpha(\omega_L)\right|}  \label{eq:redscatter}
\end{eqnarray}
\end{subequations}
Using Eq. \req{eq:bluescatter}, we see that, for $^{133}$Cs in a blue-detuned optical lattice with the reference parameters given in Section~\ref{sec:intro}, we obtain a Raman scattering rate of $\Gamma /\hbar = 2.2 \times 10^{-4}\un{s^{-1}}$, and thus an effective EPG value of $(2.2 \times 10^{-4}\un{s^{-1}}) \frac{N T_1}{n_A}$.

\subsubsection{Qubit loss and leakage}
Qubit loss errors are particularly serious, in that they can not be automatically corrected by error correcting codes. When an atom is lost from the optical lattice, or leaks into a non-qubit state, it is necessary to first detect the error before it can be corrected. The lost atom must be replaced before standard erasure error correcting codes~\cite{grassl:33} can be applied to correct the error. Detecting qubit loss requires that we have a method of detecting the presence of an atom at a given lattice site without disturbing its state.

Preskill~\cite{preskill:leakage} identified a simple circuit for performing such loss detection measurements. The circuit requires an ancilla in a known state, two applications of a CNOT or CPHASE gate, a similar number of single-qubit gates, and a measurement of the ancilla. This measurement could fail by giving an incorrect result (false positives or false negatives), or by disturbing the state of the target atom. The latter type of error could be corrected by standard error correcting codes, while the former could be minimized by repeating the measurement as necessary. Another possibility for detecting qubit loss involves the use of a cavity QED system~\cite{vala:qubitloss}.

The need for having certain atoms in the lattice reserved for use as ancillas for this scheme could be avoided by transporting an extra-lattice ancilla atom where needed through the use of optical tweezers~\cite{kuhr:2001dds}. If an atom loss was detected, this ancilla would already be on hand to serve as a replacement. A drawback of this approach is that performing such operations in parallel would require many sets of optical tweezers. In the case of most leakage errors, parallelizable methods exist for detecting ``leaked'' atoms and returning them to a qubit state.

Fortunately, qubit loss rates are very low, with storage times as long as 25 s already reported~\cite{schrader:150501}. Collisions with background gas atoms are the primary cause of loss, and so it appears that storage times can be increased further through improved vacuum systems. It is also possible that a method may be found for performing loss detection measurements in parallel, which, when coupled with a means for replacing lost atoms, would allow qubit loss errors to be handled by standard error correction techniques. Consequently, qubit loss is not likely to be the dominant source of errors in the near future, and we will not consider it further in this paper.

\subsection{Raman-based single qubit gates\label{sec:ramansinglequbit}}
Two-photon Raman transitions present an attractive option for single qubit gates because of the associated speed of qubit manipulation.  Raman-based single qubit rotations have recently been experimentally demonstrated on a time scale less than 100 ns for a single $^{87}$Rb atom trapped in an optical dipole trap ~\cite{jones:040301}.  A theoretical analysis of factors contributing to gate imperfections for a single $^{87}$Rb atom concluded that gate fidelities of $\sim 10^{-4}$ are possible~\cite{saffman:022347}.
 We analyze here the error mechanisms arising during Raman gates implemented for $^{133}$Cs atomic qubits in a blue-detuned optical lattice.  

We consider a Raman process in which the $6S_{1/2}(F=3) \leftrightarrow 6P_{1/2}$ transition is driven with strength $\Omega_1$ by 
$\epsilon_{+1}$-polarized light at a detuning of $\Delta_1$, and the $6P_{1/2} \leftrightarrow 6S_{1/2} (F=4)$ transition is driven with strength $\Omega_2$ by $\epsilon_{+1}$-polarized light at a detuning $\Delta_2$.

In the general case of a Raman-based single qubit gate with two-photon detuning $\Delta=\Delta_1 - \Delta_2 + (|\Omega_2|^2 - |\Omega_1|^2)/2(\Delta_1 + \Delta_2)$, Rabi frequency $\Omega_R\simeq\Omega_1\Omega_2^\ast / (2 \Delta_1)$, and pulse duration $t$, we have an effective off-resonance Rabi frequency $\Omega^\prime\simeq \sqrt{|\Omega_R|^2+\Delta^2}$, and the rotation is approximately described by the following matrix~\cite{saffman:022347}:
\begin{widetext}
\begin{eqnarray}
R(\Omega_R,\Delta,t) & = & \left(\begin{array}{cc}e^{i\Delta t /2} \left[\cos\left(\frac{\Omega^\prime t}{2}\right)-i\frac{\Delta}{\Omega^\prime}\sin\left(\frac{\Omega^\prime t}{2}\right) \right]& i e^{i\Delta t /2} \frac{\Omega_R^\ast}{\Omega^\prime}  \sin\left(\frac{\Omega^\prime t}{2}\right) \\ i e^{-i\Delta t /2} \frac{\Omega_R}{\Omega^\prime}  \sin\left(\frac{\Omega^\prime t}{2}\right) & e^{-i\Delta t /2} \left[\cos\left(\frac{\Omega^\prime t}{2}\right)+i\frac{\Delta}{\Omega^\prime}\sin\left(\frac{\Omega^\prime t}{2}\right) \right] \end{array} \right).\label{eq:fullrotation}
\end{eqnarray}
\end{widetext}
Unless otherwise noted, we assume zero two-photon detuning, i.e., $\Delta=0$. For the 
specific resonance case $\Delta_1 \simeq \Delta_2$, the rotation matrix is as follows:
\begin{eqnarray*}
R(\theta, \phi) & = & \left(\begin{array}{cc} \cos(\theta/2) & i e^{-i \phi} \sin(\theta/2) \\ i e^{i \phi} \sin(\theta/2) & \cos(\theta/2) \end{array} \right),
\end{eqnarray*}
with $\theta=|\Omega_R|t$ and $\phi=\arg(\Omega_R)$. 

It is necessary to define a metric for fidelity of rotation operations. We consider a qubit in an arbitrary initial state $\psi$ undergoing a rotation $R(\theta,\phi)$, and compare it to a $\pi$ pulse, $R_0(\pi,0)$. The fidelity is then given by the following relation \req{eq:gatefidelityB}:
\begin{subequations}
\begin{eqnarray}
F & = & \overline{\left| \langle \psi | R_0^\dag R | \psi \rangle \right|^2} \label{eq:gatefidelityA} \\
& = &\frac{1}{2} +  \frac{1}{6} \left(\cos 2 \phi - 2 \cos \theta \cos^2 \phi \right), \label{eq:gatefidelityB}
\end{eqnarray}
\end{subequations}
where the overline represents an average over initial states and the corresponding error is given by $P=1-F$. (Note that this definition differs from~\cite{saffman:022347} which considered the fidelity of a $\pi/2$ pulse on a specific initial state.)

\subsubsection{Neighbor atom errors\label{sec:neighbor}}
In the case of a single qubit gate performed with two orthogonal Raman lasers, an atom that is 
adjacent to the target atom and that is on the axis of one of the two lasers will experience a small undesired rotation. The effective Rabi frequency for this non-target atom is $\Omega_{R}^\prime=\Omega_R I(a,a)/I_0 $, where $I(r,z)$ is given by Eq. \req{eq:gaussian}. From Eq. \req{eq:gatefidelityA}, we can determine the fidelity error in the desired identity operations for the four neighboring non-target atoms as:
\begin{eqnarray}
P_{rn} & \simeq &  \frac{2 \pi^2}{3} \left(\frac{\Omega_R^\prime}{\Omega_R}\right)^2 \label{eq:ramannontarget} \\
& = &  \frac{2 \pi^2}{3}\left(1+\frac{a^2 \lambda_R^2}{\pi^2 w_0^4} \right)^{-2} \exp\left[-\frac{4a^2}{w_0^2}\right], \nonumber
\end{eqnarray}
where $\lambda_R\simeq894\un{nm}$ is the Raman laser wavelength

\subsubsection{Spontaneous emission\label{sec:ramanspontaneous}}
In the limit where $\Delta_1 \simeq \Delta_2$, the probability of spontaneous emission during a $\pi$ pulse is $P_{rs}\simeq \frac{\pi}{2 |\Delta_1| \tau}$, where $\tau = 34.9\un{ns}$ is the natural lifetime of the $6P_{1/2}$ state. For $\left|\Omega_R \right| << \left|\Delta_1\right|$, the detuning $\Delta_1$ is related to the Rabi frequency and Raman laser intensity $I$ by
\begin{eqnarray}
|\Omega_R| & = & (8.3\cdot 10^{12}  \un{W}^{-1}  \un{m}^2\un{s}^{-2}) \frac{I} {|\Delta_1|}; \label{eq:omR}
\end{eqnarray}
therefore the probability of spontaneous emission is
\begin{eqnarray}
P_{rs} & \simeq & \frac{\pi}{2} \frac{| \Omega_R|}{I  \tau (8.3\cdot 10^{12}  \un{W}^{-1}  \un{m}^2\un{s}^{-2}) }. \label{eq:omRPrs}
\end{eqnarray}

\subsubsection{Raman beam AC Stark Shifts\label{sec:ramanraman}}
The difference in AC Stark Shift between the logical $\ket{0}$ and $\ket{1}$ states gives rise to a phase shift $\delta \phi = t (\delta U_1 - \delta U_0)/\hbar$. For a $\pi$ pulse, $t=\pi / |\Omega_R|$, so for $\epsilon_{+1}$ polarized light we have $\delta \phi = \pi \left(\frac{\delta U_1 - \delta U_0}{\hbar \Omega_R}\right)$. In the range $50\un{GHz} < \Delta_{1}/2 \pi < 5000 \un{GHz}$, the ratio $\frac{\delta U_1 - \delta U_0}{\hbar \Omega_R} \simeq (-6 \cdot 10^{10} \un{s}^{-1}) / \Delta_{1}$. 

We now wish to calculate the variance $\mathrm{var}(\delta \phi) = \langle \delta \phi^2 \rangle - \langle \delta \phi \rangle^2$ due to atomic motion and spatial variation in the Raman beam intensity. Since, for typical parameters, the Rayleigh length of the Raman beam will be much larger than the beam waist ($z_0 \gg w_0$), we need consider only motion in the transverse direction. In the transverse direction at the beam waist, the intensity has the form $I(r,0)=I_0 e^{-2r^2/w_0^2}$, where $I_0$ is the intensity at the center, $w_0$ the beam waist, and $r$ the transverse distance from the center. 
The atomic motional states can be approximated by eigenstates of the two-dimensional harmonic trapping potential obtained by parabolic expansion of the transverse potential at the minima of the lattice potential.

For an atom in the resulting two-dimensional harmonic oscillator eigenstate $\ket{n_x,n_y}$, we calculate the variance of the phase shift using the fourth-order Taylor expansion of the Gaussian beam intensity~\footnote{Saffman and Walker in Ref.~\cite{saffman:022347} appear to consider an approximation of the form $I(r,0) \simeq I_0 / (1 + 2 r^2 / w_0^2)$, which, while generally a reasonable approximation for a Gaussian beam, is less accurate in the center region ($r<w_0/\sqrt{2}$) than the fourth-order Taylor expansion $I(r,0) \simeq I_0 (1 - 2 r^2 / w_0^2 + 2 r^4 / w_0^4)$.}, 
$I(r,0) \simeq I_0 (1 - 2 r^2 / w_0^2 + 2 r^4 / w_0^4)$. This results in the variance $\mathrm{var}_{n_x,n_y} (\delta \phi)$
\begin{eqnarray}
\mathrm{var}_{n_x,n_y}(\delta \phi) & \simeq & \frac{\hbar^2 a^2}{\pi^2 m U_L w_0^4} \left(\frac{6 \cdot 10^{10} \un{s}^{-1}}{\Delta_{1}} \right)^2  \label{eq:phasevariance} \\
& \times & \left[\left( n_x^2 + n_x + 1\right)+\left( n_y^2 + n_y + 1\right) \right]. \nonumber
\end{eqnarray}
(In studying the temperature dependence of this effect, the reader may find it helpful to make the approximation $k_BT/2 \simeq \hbar \omega (n_x+1/2) \simeq  \hbar \omega (n_y+1/2)$.) From the expression for fidelity of a $\pi$ pulse, Eq. \req{eq:gatefidelityB}, we see that the expected error probability will be $P_{ra}=\frac{2}{3} \mathrm{var}_{n_x,n_y} (\delta \phi)$.

\subsubsection{Atomic motion\label{sec:atomicmotion}}
In addition to the effects discussed above, atomic motion will introduce noise through variation in the effective pulse area, $|\Omega_R| t$, and variation in the two-photon detuning, $\Delta$. The former effect is simply a result of atomic motion across the Gaussian profile of the Raman beams, and has a similar form to the result calculated in the previous section, Eq.  \req{eq:phasevariance}. For a $\pi/2$ pulse, we obtain the following result for the variance:
\begin{eqnarray}
\lefteqn{\mathrm{var}_{n_x,n_y}(|\Omega_R| t)  \simeq}  \label{eq:pulsevariance} \\
 &  & \frac{\hbar^2 a^2}{\pi^2 m U_L w_0^4} \left(\frac{\pi}{2}\right)^2\left[\left( n_x^2 + n_x + 1\right)+\left( n_y^2 + n_y + 1\right) \right]  \nonumber
\end{eqnarray}
This variation in pulse area will then result in an error 

$P_{rpa}=\frac{1}{6}\mathrm{var}_{n_x,n_y}(|\Omega_R| t)$ for a $\pi/2$ gate.

Doppler shifts of the Raman beams will cause variation in the two-photon detuning $\Delta$. Unlike the isolated two-site dipole trap situation considered by Saffman and Walker~\cite{saffman:022347}, our system involves a 3D lattice and thus does not allow for a convenient first-order Doppler-free Raman laser configuration. We thus expect to see significant variation in the two-photon detuning due to atomic motion-induced Doppler shifts, as described by the following relation:
\begin{eqnarray}
\mathrm{var}_{n_x,n_y}(\Delta) & \simeq & \left(\frac{2 \pi}{\lambda_R}\right)^2 \left\langle v^2 \right\rangle \label{eq:deltavariation} \\
 & = & \simeq \left(\frac{2 \pi}{\lambda_R}\right)^2 \frac{\hbar \omega_\tau}{m}\left(n_x + n_y + 1\right)
\end{eqnarray}
From the general expression for the rotation matrix, Eq.  \req{eq:fullrotation}, we determine that this variation will result in a fidelity error $P_{rm} \simeq \frac{8 - 4 \pi + \pi^2}{24} \Omega_R^{-2} \mathrm{var}_{n_x,n_y}(\Delta) \simeq 0.22\  \Omega_R^{-2} \mathrm{var}_{n_x,n_y}(\Delta) $

\subsubsection{Polarization effects\label{sec:polarization}}
The Raman beams used to perform the single-qubit gate have a Gaussian profile. This means that, even at the beam waist, the beam will have a small component 
of polarization other than the desired $\epsilon_{+1}$, according to~\cite{saffman:022347}:
\begin{eqnarray}
E(x,y,0) & = & \frac{E_0}{2}\left(\epsilon_{+1} + \frac{(y -ix)}{z_R}\epsilon_{0}\right) e^{-\frac{(x^2+y^2)}{w_0^2}}+\mathrm{c.c.},
\label{eq:E_otherpol}
\end{eqnarray}
where $z_R = \pi w_0^2 / 2$ is the Rayleigh length.
This extraneous polarization can result in leakage errors by causing transitions to states outside the computational basis. To estimate the probability of such errors, we determine the relative magnitude of the second term 
above, which corresponds to undesired polarization ``seen'' by a target atom in a vibrational state $\ket{n_x,n_y}$.  Eq.~(\ref{eq:E_otherpol}) suggests that this can be estimated by the ratio of the spatial extent of the atom to the Rayleigh length, i.e., by 
\begin{eqnarray}
|\mathcal{P}_0|_{\mathrm{rel}} & \simeq & \sqrt{\left \langle r^2 \right \rangle } / z_R \label{eq:polmatel} \\
& = & \frac{\lambda_R }{\pi w_0^2} \sqrt{\frac{\hbar}{m \omega_\tau} \left(n_x + n_y +1\right)} , \nonumber
\end{eqnarray}
where $\omega_\tau =\frac{\pi}{a} \sqrt{2 U_L / m}$ is the characteristic trapping frequency in the harmonic approximation.
There are four possible non-basis states into which the qubit could leak: $\ket{F=3,m_F=\pm1}, \ket{F=4,m_F=\pm1}$. Since the matrix elements for the unwanted transitions are comparable to those for the desired transitions~\cite{steck:cesium}, and the corresponding Rabi frequencies are smaller by a factor of $|\mathcal{P}_0|_{\mathrm{rel}}$, the probability of leakage into any particular state for a $\pi$ gate can be approximated by
 $\sin^2\left( |\mathcal{P}_0|_{\mathrm{rel}} \pi / 2\right)^2 \simeq \left( |\mathcal{P}_0|_{\mathrm{rel}} \pi / 2\right)^2 $. The total leakage probability in the 
 motional ground state, $n_x=n_y=0$,  is then four times that quantity, i.e.:
\begin{eqnarray}
P_{rp} &   \simeq & \frac{\hbar \lambda_R^2 a}{\pi w_0^4 } \frac{1}{\sqrt{2 mU_L}}.
\end{eqnarray}

\subsubsection{Laser intensity noise and line-width\label{sec:lasernoise}}
Noise in the Raman lasers affects the fidelity of the gate. If the relative intensity fluctuation is $\delta I / I$, then by Taylor-expanding eqn. \req{eq:gatefidelityB} with $\theta\rightarrow \theta_0 + \delta \theta$, we see this will result in an initial state-averaged fidelity error of 
$P=(1/6)(\pi \delta I / I)^2$ for a $\pi$ gate and $P=(1/6)\left(\frac{\pi}{2} \delta I / I \right)^2$ a $\pi/2$ gate. 

Even if the Raman lasers are actively stabilized, shot noise provides a lower bound on relative intensity fluctuations. The fluctuation due to shot noise ~\cite{wineland:ions} is given by 
\begin{eqnarray}
\frac{\delta I}{I} & \geq &\left(\frac{4 \hbar \omega_R}{\eta P_R t_\pi}\right)^{1/2}\!\!\!\!\!\!, \label{eq:shotnoise}
\end{eqnarray}
where $\eta$ is the quantum efficiency of the detector used in the stabilization circuit, $\omega_R$ is the frequency of the Raman laser, and $P_R$ is the laser power. If we assume $\eta = 0.5$ and that our stabilization circuit reaches the lower bound, then the minimum fidelity error is
\begin{eqnarray}
P_{rl} & = & \frac{4 \pi^2 \hbar \omega_R}{3 P_R t_\pi}. \label{eq:shotfidelity}
\end{eqnarray}
For detunings much smaller than the absolute optical frequency, we can use $\omega_R \simeq 2 \pi \cdot 3.5 \cdot 10^{14} \un{s}^{-1}$ (i.e., the Raman transition frequency). For a Gaussian beam of power $P_R /2$, the intensity $I$ at the waist is related to the power as $I= P_R / \pi w_{0}^2$, while $t_\pi = \pi / \Omega_R$, with $\Omega_R$ given by \req{eq:omR}. This results in the following estimate for laser intensity fluctuation-induced error:
\begin{eqnarray}
P_{rl} & \simeq & (2.6 \cdot 10^{-6} \un{m}^2 \un{s}^{-1}) \frac{1}{|\Delta_1| w_{0}^2}. \label{eq:shotfidelity2}
\end{eqnarray}

\subsection{Microwave-based single qubit gates\label{sec:microwavesinglequbit}}
Site-specific single-qubit gate operation in 3D lattices can also be achieved through the use of a far off-resonance addressing laser focused on a single lattice site, combined with pulsed global microwave fields~\cite{vala:0307085,scheunemann:051801, scheunemann:lattice, zhang:042316}. In order to address a single atom, a Gaussian beam with waist substantially smaller than the lattice spacing is used, which results in the target atom seeing a much greater field than any neighboring atom. The intensity of a Gaussian beam is described by 
\begin{eqnarray}
I(r,z) & = & I_0 \frac{w^2(z)}{w^2_0} \exp{\left(-2 r^2 / w^2(z)\right)} \label{eq:gaussian}
\end{eqnarray}
where $w_0$ is the beam waist, $I_0$ the intensity at the center of the waist, $z_0=\pi w_0^2 / \lambda$ is the Rayleigh length, and $w(z)=w_0 \sqrt{1+ \frac{z^2}{z_0^2}}$ the beam width as a function of the axial coordinate $z$.

\begin{figure}[pt] \centering
\includegraphics[width=2.7 in,  keepaspectratio=true]{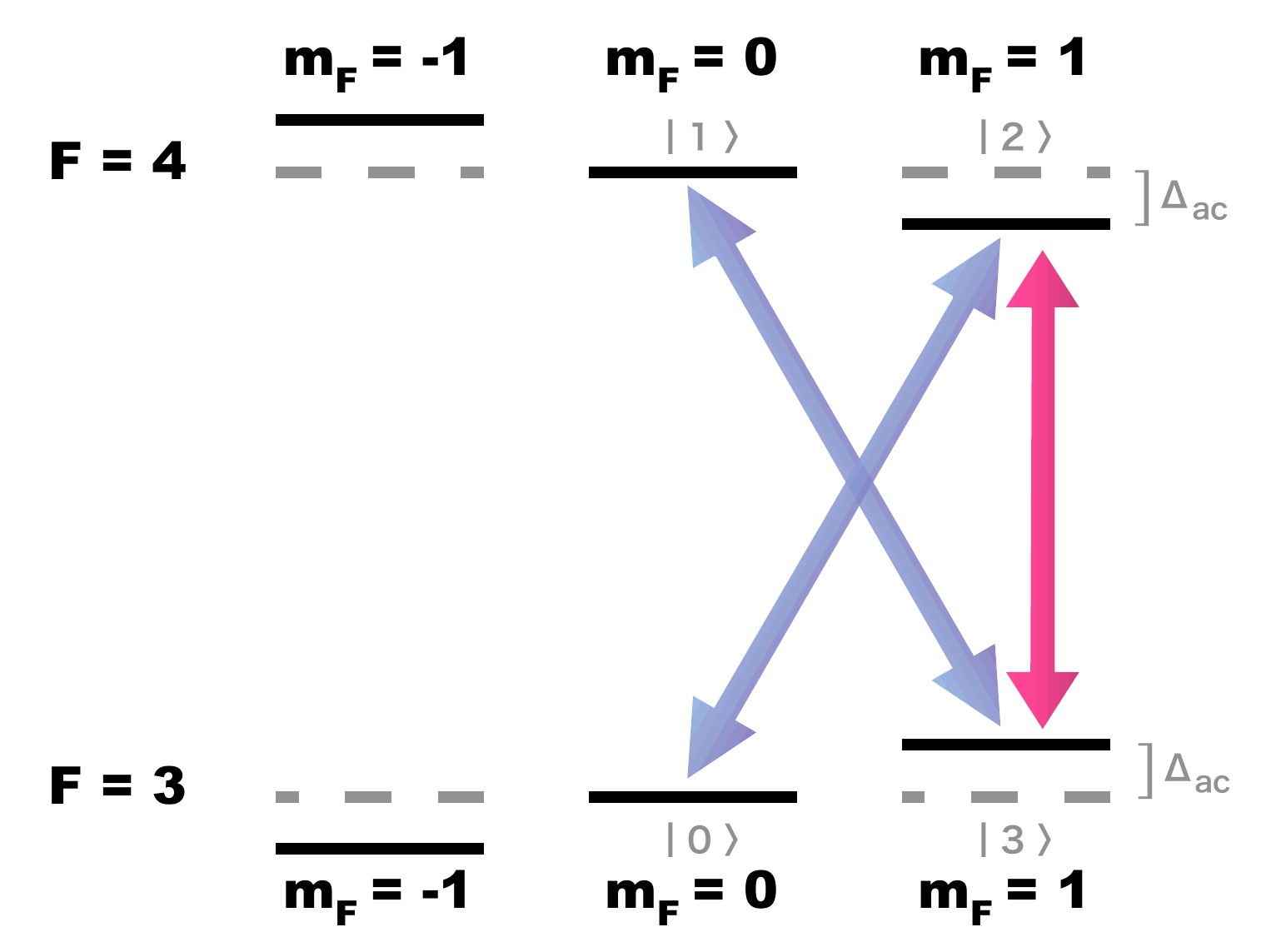}
\caption{\label{fig:gate} (Color online) Schematics of a single qubit flip operation. A focused far off-resonant addressing laser produces an AC Stark Shift $\Delta_{ac}$ of the $\ket{F=4, m_F=1}$ and $\ket{F=3, m_F=1}$ levels in a single target atom.   The qubit levels are the $m_F=0$ levels, labeled $\ket{0}$ and $\ket{1}$.  Levels $\ket{2}$ and $\ket{3}$ are auxiliary levels involved in the single qubit gate. (Most of the hyperfine sub-levels not involved in the gate are not shown here.) Three global microwave pulses tuned to the transitions $\ket{0} \leftrightarrow \ket{2}$, $\ket{1} \leftrightarrow \ket{3}$, and $\ket{2} \leftrightarrow \ket{3}$ drive these transitions in the target atom with transition strengths $\Omega_2$, $\Omega_2$, and $\Omega_1$.  Other atoms are not affected to first order (see below) since their transitions are not resonant.  Our simulations indicate that best performance is achieved by simultaneously driving all transitions, with the relative transition strengths chosen such that $\Omega_2 = \sqrt{3}/2 \Omega_1$.}
\end{figure}

The addressing beam causes an AC Stark Shift $\Delta_{ac}$ of the various levels of the target atom. Here we consider the scheme for $^{133}$Cs that is outlined in Figure~\ref{fig:gate}. By choosing the ``magic wavelength'', $\lambda_M$, for the addressing beam (for $^{133}$Cs, 
$\lambda_M\simeq 880\un{nm}$), the $\ket{2}$ and $\ket{3}$ auxiliary levels receive AC Stark Shifts of equal magnitude but opposite sign, while the qubit levels $\ket{0}$ and $\ket{1}$ are unaffected. This allows the $\ket{0} \leftrightarrow \ket{2}$, $\ket{2} \leftrightarrow \ket{3}$, and $\ket{1} \leftrightarrow \ket{3}$ transitions to be driven by global microwave pulses that are resonant only for the target atom. Alternatively, a collimated beam could be used to address an entire row of atoms (provided the row was not much longer than the Rayleigh length $\pi w_0^2 / \lambda_M$), allowing $n_A = N^{1/3}$ identical operations to be performed in parallel. However, since the beam waist $w_0$ must be smaller than the lattice spacing $a$, this implies that $n_A < \pi a / \lambda_M$ and consequently only a relatively small number of atoms can be addressed simultaneously using this method. We discuss this limitation further in the next subsection.

We have developed a software package, quantum simulation software (QSIMS)~\cite{QSIMS}, for simulating the quantum dynamics of one and two-qubit gates in this and other systems. Using QSIMS, we discretize the spatial wavefunction of the atom on a grid, with a separate grid representing each possible internal state of the atom. Quantum dynamics are simulated by applying the Schr\"odinger propagator, expanded in Chebychev polynomials~\cite{kosloff:propagation}. The kinetic portion of the Hamiltonian is applied by means of a Fourier transform of the discretized wavefunction from the position basis to the momentum basis. Transitions between levels are treated with a dressed state approach~\cite{cohentannoudji:atomphoton}.
Although QSIMS is capable of simulating three spatial dimensions, the symmetry of the system and the near-separability of the lattice potential make it reasonable in most cases to perform simulations in only one or two spatial dimensions. This results in a significant speed-up, since the run time of the simulations is $\mathcal{O}(N \log N)$ for a grid of $N$ points. 

We simulate the microwave pulse-based single qubit $\pi$ gate with QSIMS, using the parameters $a = 5\un{\mu m}$, $U_L=200 \un{\mu K}$, $\Delta_{ac} = 0.2 \un{MHz}$, $w_0 = 1.2 \un{\mu m}$, and $T_1 = 76 \un{\mu s}$ (with microwave pulse intensity $\Omega_1 = 41341 \un{s}^{-1}$ chosen appropriately to achieve this gate time).  The atom is assumed to be initially in the $\ket{0}$ qubit state and in the motional ground state, so the final state of the gate correspond to $\ket{1}$ and the motional ground state.  We investigated various different versions of the gate with QSIMS and found that the best performance is achieved by simultaneously driving all three transitions $\ket{0} \leftrightarrow \ket{2}$, $\ket{1} \leftrightarrow \ket{3}$, and $\ket{2} \leftrightarrow \ket{3}$, with pulse intensity chosen such that the strength $\Omega_2$ of the first two transitions is $\sqrt{3}/2 \Omega_1$, where $\Omega_1$ is the strength of the $\ket{2} \leftrightarrow \ket{3}$ transition. 

\subsubsection{Off-resonant transitions\label{sec:offresonant}}
Since the microwave pulses proposed to perform these single qubit operations are applied globally, i.e., to the entire lattice, there is a small probability that such a pulse will cause a given non-target atom to undergo a non-resonant transition. We can minimize this probability by carefully tuning the gate parameters such that the pulse ends with non-target atoms in a local minima of their Rabi cycles, and by making the detuning, $\Delta$, large compared with the Rabi coupling $\Omega_1$. We are limited in our ability to do the former by our pulse timing resolution, $\delta_T$, the uncertainty in pulse length $T_1$. 

Most qubits in the lattice will be far from the target qubit so that $I(r,z)$ is small (Eq.~(\ref{eq:gaussian})), and thus will not experience any Stark shift due to the addressing beam. The $\ket{0}\leftrightarrow\ket{2}$ and $\ket{1}\leftrightarrow\ket{3}$ transitions are detuned from these unshifted qubit transitions by $\Delta_{ac}$, while the $\ket{0}\leftrightarrow\ket{1}$ transition is detuned by $2 \Delta_{ac}$. Since the probability $P$ of transition for any given atom is small, we can treat the transition amplitudes as independent, and calculate each transition probability using the Rabi formula, Eq.~\req{eq:rabicalcA}, assuming that the coupling and pulse time are chosen such that $\sin{\left( \sqrt{\Omega_1^2 + \Delta^2} \frac{T_1^\prime}{2} \right)} = 0$ and $T_1^\prime \simeq \pi / \Omega_1$, where $T_1^\prime \simeq T_1/3$ is the time required for one ``leg'' of the single qubit gate (see above and Figure~\ref{fig:gate}). Since we wish to minimize $\Delta_{ac}$ for the purposes of reducing other types of errors discussed below, we note that the smallest value of $\Delta$ for which the former condition is satisfied is $\Delta = \sqrt{3} \Omega_1$.  This results in the off-resonant transition probability estimates 
\begin{subequations}
\begin{eqnarray}
P_{mo} & = & \frac{\Omega_1^2}{\Omega_1^2 + \Delta^2} \sin^2{\left( \sqrt{\Omega_1^2 + \Delta^2} \frac{T_1^\prime \pm \delta_T}{2} \right)} \label{eq:rabicalcA} \\
& \simeq & \left(\frac{\pi}{2} \frac{\delta_{T_1}}{T_1^\prime} \right)^2. \label{eq:rabicalcB}
\end{eqnarray}
\end{subequations}
With these estimates we can now ask, what is the corresponding EPG due to off-resonant transitions of all non-target atoms? If we can simultaneously address an entire row of 
$n_A=N^{1/3}$ atoms in a lattice of $N$ atoms, the EPG is $N^{2/3} P_{mo}$. Unfortunately, this is challenging in even a modestly-sized lattice, as the intensity of the beam is inhomogeneous along the beam axis, on a length scale set by the Rayleigh length $z_R$.  By combining equations \req{eq:gaussian} and \req{eq:rabicalcA}, with $\Delta= \sqrt{3} \Omega_1\left(1-\frac{1}{1+z^2/z_R^2}\right)$, we estimate that the error for an atom at distance $z$ from the beam waist along the beam axis would be approximately $1-P_{mo}(z) \simeq \frac{9 \pi^2}{16} \frac{z^8}{z_R^8}$.  For typical parameter values, this limits us to addressing just a few lattice sites with a single beam before the error becomes large. In fact for our example parameters, the Rayleigh length is $z_R=5.1\ \mu \mathrm{m}$, and the error becomes of order unity at just one lattice site away from the beam waist.

If we cannot simultaneously address entire rows with a single addressing beam, the effective EPG will scale as $N \left(\frac{\pi}{2} \frac{\delta_{T_1}}{T_1^\prime} \right)^2$.  Although such scaling of an error mechanism would preclude scalable fault-tolerant quantum computation for an arbitrarily-large system, in practice it should not prove very restrictive for lattices with moderate numbers of qubits. For example, for microwave pulses of the appropriate frequency, $\delta_T$ can be on the order of one cycle, or $10^{-10}$ s. For a single-qubit gate time of $T_1=10 \mu \mathrm{s}$, this means a lattice of $10^{5}$ atoms could have an EPG of less than $10^{-5}$ due to off-resonant transitions. Scaling implications are discussed further in Section \ref{sec:analysis}. 

\subsubsection{Addressing beam-induced heating\label{sec:addressheat}}
\begin{figure}[pt] \centering
\includegraphics[width=3.4 in,  keepaspectratio=true]{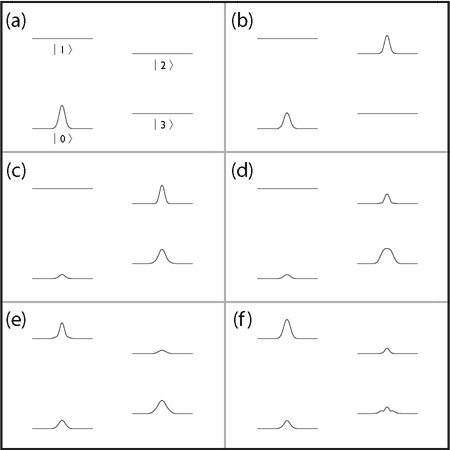}
\caption{\label{fig:qsims} A series of ``snapshots'' of the atomic spatial and internal state wavefunction for  an atom in a 1D trapping potential undergoing a microwave pulse-based single qubit $\pi$ rotation gate, as simulated by QSIMS. The four relevant internal atomic states are labeled $\ket{0}$ through $\ket{3}$, as in Fig. \ref{fig:gate}, and the curves represent the wavefunction for the atom's center-of-mass coordinate. Simulation parameters have been chosen to exaggerate the gate errors so that they are visible in this figure. In (a), we see the initial state of the atom is the internal state $\ket{0}$ ($F=3$, $m_F=0$), with a Gaussian center-of-mass wavefunction. In (b) and (c), we see transitions to the auxiliary levels $\ket{2}$ and $\ket{3}$. In (d) and (e), vibrationally-excited states become visible, particularly for the component of the total wavefunction in the state $\ket{3}$. In (f), we see the final state with a significant portion of the wavefunction not in the desired $\ket{1}$ state, showing instead noticeable entanglement between spatial and internal degrees of freedom (e.g., the ``bumpy'' shape of the wavefunction component for the $\ket{3}$ level).}
\end{figure}

The far off-resonant Gaussian addressing beam used to perform site-selective single qubit gates  contributes harmonic trapping and anti-trapping 
potential terms for the $\ket{2}$ and $\ket{3}$ states, respectively, and also adds additional state-dependent anharmonic terms to the potential experienced by the atom. These anharmonic terms can generate entanglement between motional and internal degrees of freedom, as well as ``heating'' the atom to higher motional states. Perturbation theory shows that the most significant undesirable effect is due to the difference between the harmonic components of the trapping potentials experienced by atoms in the auxiliary states $\ket{2}$ and $\ket{3}$, relative to those experienced in qubit states $\ket{0}$ and $\ket{1}$.

The overlap between the vibrational ground state of the $\ket{0}$ level and the first even vibrationally-excited state of the $\ket{2}$ level is approximately $\xi_{T} =\frac{\sqrt{2}}{2} \frac{\hbar \Delta_{ac}}{m \omega_\tau^2 w_0^2}$, where $\omega_\tau$ is the 
characteristic trapping frequency defined above and $w_0$ is the addressing beam waist. From this, we can calculate the probability of transition to the first even vibrationally-excited state, using 
the fact that $\omega_\tau \gg \xi_{T} \Omega_1$ and $T_1 \simeq \pi / \Omega_1$:
\begin{eqnarray}
P_{mh} & = & \frac{\Omega_1^2 \xi_{T}^2}{\Omega_1^2 \xi_{T}^2 + 4 \omega_\tau^2} \sin^2{\left( \sqrt{\Omega_1^2 \xi_{T}^2 + 4 \omega_\tau^2} \frac{T_1}{2} \right)} \label{eq:addrbeamheat} \\
& \simeq & \left(\frac{\Omega_1 \xi_{T}}{2 \omega_\tau} \right)^2  \nonumber \\
& = & \frac{1}{64 \pi^4} \frac{\hbar^2 \Delta_{ac}^2 m a^6}{T_1^2 U_L^3 w_0^4} \nonumber
\end{eqnarray}
Here we have simplified Eq.~\req{eq:addrbeamheat}  
by assuming that, since $\omega_\tau$ is of the same order as $\Omega_1$, the $\sin^2 \left( \ldots \right)$ term will be of order unity.

We have tested these  
perturbative predictions by simulating  the single-qubit $\pi$ gate with QSIMS as described above, with initial condition $\ket{0}$.  The simulations show that the probability that the atom will not be in the motional ground state and desired qubit state after completion of the $\pi$ gate
 is $1 \times 10^{-6}$. This is consistent with what is expected from application of Eq. \req{eq:addrbeamheat} with our simulation parameters.

A small amount of heating does not in itself destroy the qubit state. However, as the vibrational energy of the atoms increases, the probability of other types of errors increases, and so the atoms will periodically need to be re-cooled to the ground state. Optical cooling can be done directly if the qubit state were first transferred to a different location, or could potentially be done with the qubit ``in place'' through a mechanism such as sympathetic cooling. Analysis of such re-cooling mechanisms is beyond the scope of this paper.

\subsubsection{Addressing beam-induced Raman Scattering\label{sec:addressscatter}}
We now determine the Raman scattering rate for the target atom during a single-qubit 
microwave gate (Rayleigh scattering at this wavelength is negligible). Using Eq. \req{eq:scatterrate}, we can find the scattering rate in terms of $E^2$. 
For $\epsilon_+$ polarized light at the magic wavelength $\lambda_M$, calculation using Eq. \req{eq:polarizability} yields a polarizability of $|\alpha^{(-)}(880\un{nm})|=2.5 \times 10^{-38}\ \mathrm{C^2\ m^2\ J^{-1}}$ for $^{133}$Cs in the \ket{F=3, m_F=1} or \ket{F=4, m_F=1} state (note that the polarizabilities for these states have opposite signs and that the polarizabilities for the $m_F=0$ states are essentially zero at $\lambda_M$). Since the AC Stark shifts of the target atom \ket{F=3, m_F=1} and \ket{F=4, m_F=1} levels are given by $\hbar \Delta_{ac} = -  \frac{E^2_0}{4}\alpha(\lambda_M)$, we can express the scattering rate in terms of $\Delta_{ac}$ as follows:
\begin{eqnarray}
\Gamma / \hbar & = & 880\un{nm} \frac{\epsilon_0 \Delta_{ac}}{\pi \left| \alpha^{(-)}(880\un{nm}) \right|} \sigma(880\un{nm}) \label{eq:abscatter} \\
& \simeq & 3.4 \times 10^{-6}\ \Delta_{ac}. \nonumber
\end{eqnarray}
We then obtain the corresponding Raman scattering error per gate, $P_{ms}$, by multiplying the scattering rate $\Gamma / \hbar$ 
of Eq. (\ref{eq:abscatter}) by the single-qubit gate time $T_1$.

\subsubsection{Addressing beam position error\label{sec:position}}

If the addressing beam is off-target by an amount $\delta_x$, Taylor expansion of \req{eq:gaussian} shows that the energy of the $\ket{1}$ and $\ket{2}$ levels will be shifted by an amount $2 \hbar \Delta_{ac} \frac{\delta_x^2}{w_0^2}$. This decreases the $\ket{0} \rightarrow \ket{1}$ transition probability 
according to:
\begin{eqnarray}
1-P_{mpt} & = & \frac{\Omega_1^2}{\Omega_1^2 + 4 \Delta_{ac}^2 \frac{\delta_x^4}{w_0^4}} \sin^2 \left(\sqrt{\Omega_1^2 + 4 \Delta_{ac}^2 \frac{\delta_x^4}{w_0^4}} \frac{T_1}{2} \right) \nonumber \\
& \simeq & 1- \frac{4}{\pi^2}\Delta_{ac}^2 T_1^2  \frac{\delta_x^4}{w_0^4}.  \label{eq:poserror}
\end{eqnarray}

There is also an effect due to the perturbation of the eigenstates of the $\ket{2}$ and $\ket{3}$ states. The matrix element between the unperturbed harmonic oscillator ground state and the perturbed first excited state is $\xi_{\delta x} = 4 \frac{ \Delta_{ac}}{\omega_\tau} \frac{\delta_x}{w_0^2} \sqrt{\frac{\hbar}{2 m \omega_\tau}}$. Assuming $\xi_{\delta x} \Omega_1 \ll \omega_\tau$, the probability of exciting to a higher motional state during the $\ket{0} \leftrightarrow \ket{2}$ or $\ket{1} \leftrightarrow \ket{3}$ transitions is then:
\begin{eqnarray}
P_{mph} & = & \frac{\Omega_1^2 \xi_{\delta x} ^2}{\Omega_1^2 \xi_{\delta x} ^2 +  \omega_\tau^2} \sin^2 \left(\sqrt{\xi_{\delta x} ^2 \Omega_1^2 +  \omega_\tau^2} \frac{T_1}{2}\right) \label{eq:poswave} \\
& < & \left( \frac{\Omega_1 \xi_{\delta x}}{\omega_\tau}  \right)^2 \nonumber \\
& =& \frac{\sqrt{2}}{\pi^3} \frac{\hbar \Delta_{ac}^2 \delta_x^2 a^5 m^{3/2}}{T_1^2 U_L^{5/2}  w_0^4} \nonumber
\end{eqnarray}
For typical parameter values, this second effect is of greater significance and we will neglect the former in comparison with this. Note also that we have calculated the error only for one ``leg'' (i.e., transition) out of the three that compose the gate, and that the total error for the gate may be greater. 

We have simulated the one-qubit microwave pulse-based $\pi$ gate between $\ket{0}$ and $\ket{1}$
using QSIMS with an addressing beam position error $\delta_x = 0.01 \un{\mu m}$.  We find that on completion of the gate, the probability that the atom will not be in the motional ground state and desired 
qubit state $\ket{1}$, is $2 \times 10^{-5}$. Using Eq. \req{eq:poswave} with the same parameters as this simulation yields a value $1.3 \times 10^{-6}$. This is an estimate of the error in the $\ket{0} \rightarrow \ket{2}$ and $\ket{3} \rightarrow \ket{1}$ legs of the gate. For the $\ket{2} \rightarrow \ket{3}$ leg, we replace $\delta_x \rightarrow 2 \delta x$ (because the perturbation due to the addressing beam has opposite sign for the $\ket{2}$ state versus the $\ket{3}$ state), to obtain an error of $7.8 \times 10^{-6}$. Summing these three errors, we obtain an overall error estimate of $7.8 \times 10^{-6}$ for the complete gate, which is within a factor of 3 of the value obtained from the simulation.

\subsection{Two-qubit gates\label{sec:twoqubit}}
In this section, we make a qualitative comparison of the two-qubit gate techniques most commonly mentioned in the literature, with an emphasis on analysis of their different implications for scalability. Most schemes either involve stationary qubits and long-range interactions, or movable qubits and short-range collisional interactions between qubits.

\subsubsection{Long-range interaction-based gates}

As with single qubit gates, our choice of a large lattice spacing allows for two-qubit operations to be performed in a site-specific manner. If the atoms are to remain stationary, a long-range interaction is required to perform a two-qubit gate. Furthermore, the interaction must somehow be controllable. Dipole-dipole interactions between atoms excited into Rydberg states are a promising candidate, and many variations on this idea have been proposed~\cite{jaksch:2208,saffman:022347,lukin:dipole,protsenko:052301,safronova:040303,ryabtsev:0402006,cozzini:0511118,brennen:062309}.

In one version of the Rydberg gate~\cite{jaksch:2208}, two nearby atoms are conditionally excited into Rydberg states via a coherent process~\cite{cubel:023405}. If both atoms were initially in the $\ket{0}$ state, they are both excited into the Rydberg state where they interact via a dipole-dipole interaction, resulting in accumulation of a phase on the two-qubit state. They are then de-excited via another series of Raman pulses. Since the interactional phase accumulation occurs only in the case where both atoms are initially in the $\ket{0}$ state, this is effectively a CPHASE gate.
Estimates of both the speed and the maximum possible fidelity for Rydberg gates are reasonably promising~\cite{saffman:022347,jaksch:2208}, with some putting the error rates achievable on the order of $10^{-3}$ to $10^{-4}$~\cite{cozzini:0511118}.
However, because of the inherent long-range nature of the interactions in this gate, when running these gates between multiple pairs of qubits in parallel, it is essential to take into account the effects of cross-talk between different qubit pairs, i.e., the dipole-dipole interactions between qubits from different pairs.
To make a rough estimate of the degree of parallelization possible in the presence of such cross-talk, consider a three-dimensional lattice of atoms. Suppose an external static electric field is applied to induce a ``permanent'' dipole moment in the Rydberg-state atoms~\cite{jaksch:2208}: the level shift due to the resulting dipole-dipole interaction falls off as $1/R^3$.  The fidelity error due to cross-talk will therefore scale roughly as $1/R^6$, where $R$ is the distance between different pairs of atoms that are simultaneously involved in Rydberg gate operations.  We take the fidelity of a Rydberg gate performed between atoms one lattice spacing apart as unity, for reference. Then,
taking a value of  $10^{-6}$ for the fault tolerance threshold for gate errors (a value intermediate between different estimates of the threshold for the case of local gates~\cite{szkopek:0411111,svore:0604090}), this implies that the two-qubit gate cannot simultaneously be performed on multiple pairs of atoms within approximately ten lattice sites of each other. In a three-dimensional lattice, this geometric constraint limits us to simultaneously performing 
approximately one two-qubit gate per several hundred atoms. This in turn implies that the storage error rate during the two-qubit gate time would have to be about two orders of magnitude lower than this fault tolerance threshold (i.e., $\sim 10^{-8}$ for the above example), to avoid accumulating additional idle time errors on the qubits not involved in the gates.

It is possible to mitigate this cross-talk limitation by using an interaction with more limited range, such as the van der Waals interaction present between Rydberg atoms when there is no hybridizing static electric field
~\cite{gallagher:rydberg}. In general, interaction strengths for the zero external field case scale as $\mathcal{O}(n^\ast / R^6)$ (as is typical for van der Waals-type interactions), although there are exceptions~\cite{reinhard:rydberg}. This implies that cross talk errors would scale as $1/R^{12}$, which allows for roughly one simultaneous two-qubit gate per three dozen atoms. This imposes more modest constraints on storage error rates---the storage error rate per two qubit gate time would then only need to be about one order of magnitude below the threshold value of $10^{-6}$.

\subsubsection{Collisional gates}
Another method of avoiding the limitations of long-range interactions is to bring pairs of atoms close together and use short-range interactions or collisions to implement two-qubit gates. A variety of such gates have been proposed and analyzed~\cite{jaksch:1975,calarco:022304,stock:183201,mandel:2003,vager:022325,sebby-Strabley:0602103,joo:0601100,stock:0509211,idziaszek:0604205}. 

We note that, in the context of the system discussed in this paper, it would first be necessary to transfer atoms from the qubit states $\ket{0}=\ket{F=3,m_F=0},\ \ket{1}=\ket{F=4,m_F=0}$ to states that experience a state-selective trapping potential (i.e., states with non-zero $m_F$), so as to allow pairs of atoms to be translated towards each other and brought together. For example, for lattice light with $\lambda=800\ \mathrm{nm}$ and a particular circular polarization, the states $\ket{F=3,m_F=1}$ and $\ket{F=3,m_F=-1}$ have polarizabilities that differ by more than 8 \%. Specific atoms can be transferred into these states by microwave or Raman transitions, analogous to the single qubit gates discussed earlier. Alternatively, an entire plane of atoms can be transferred simultaneously using the microwave pulse method if the addressing beam is replaced by an inhomogeneous magnetic field. Once the atoms are in the appropriate states, the atoms can be selectively moved by changing the relative polarization of one of the lattice beams, as described by Vala \etal~\cite{vala:0307085} and Weiss \etal~\cite{weiss:040302}, allowing the atoms to be brought together to perform a gate~\cite{jaksch:1975}.

Due to the necessity of physically moving atoms around the lattice, collisional gates are likely to be much slower than long-range interaction-based gates. Even in the case of a ``fast approach'', where the translation of the atoms is not adiabatic, gate times are still one to two orders of magnitude slower than the characteristic trap period of the lattice site's potential well~\cite{vager:022325}. Estimates of the maximum fidelity possible with collisional gates also tend to be lower than maximum fidelity estimates for Rydberg gates~\cite{calarco:022304}.

Despite these drawbacks, collisional gates appear more suited to large-scale quantum computation than long-range interaction-based gates. Collisional gates can easily be performed on a massively parallel scale, particularly when used in the context of cluster-state quantum computing~\cite{raussendorf:5188,nielsen:0405134}. 
In cluster-state (also known as ``one-way'') quantum computing, two qubit gates are used only in the initial preparation of a large entangled ``cluster'' state. The computation itself is then effected via single-qubit measurements in a variety of bases, or equivalently, single qubit rotations followed by measurement in a particular basis. One recent scheme for cluster-state quantum computing offers fault tolerance thresholds as high as $10^{-2}$ for local depolarizing
errors and $10^{-3}$ if there are also errors in preparation, gates, storage, and measurement~\cite{raussendorf:0510135}. One can easily imagine building such a cluster state with atoms in a three-dimensional lattice by performing collisional gates in parallel on entire planes of atoms at a time.

\section{Analysis\label{sec:analysis}}

\begin{table*}[ht]
\begin{ruledtabular}
\begin{tabular}{lccc}
\textbf{Source} & \textbf{Section} & \textbf{EPG} \\
\hline

Raman scattering (blue-detuned lattice) & \ref{sec:latticescatter} & $T_1 \frac{N}{n_A}\frac{\pi c \epsilon_0}{a \omega_L}  \sqrt{\frac{U_L(\omega_L,E_0^2)}{2 m}} \frac{\sigma(\omega_L)}{\left| \alpha(\omega_L)\right|} $ \\
Raman scattering (red-detuned lattice) & \ref{sec:latticescatter} & $T_1 \frac{N}{n_A} \frac{2 c \epsilon_0}{\hbar \omega_L} U_L(\omega_L,E_0^2)  \frac{\sigma(\omega_L)}{\left| \alpha(\omega_L)\right|}$ \\
\hline
Neighbor atom errors (Raman gate): P$_{rn}$ & \ref{sec:neighbor} & $  \frac{2 \pi^2}{3}\left(1+\frac{a^2 \lambda_R^2}{\pi^2 w_0^4} \right)^{-2} \exp\left[-\frac{4a^2}{w_0^2}\right] $ \\
Spontaneous emission (Raman gate): P$_{rs}$ & \ref{sec:ramanspontaneous} & $ \frac{\pi}{2 |\Delta_1| \tau} $ \\
AC Stark Shifts (Raman gate): P$_{ra}$ & \ref{sec:ramanraman} & $\frac{4}{3 \pi^2} \frac{\hbar^2 a^2}{mw_0^4 U_L} \left(\frac{6 \cdot 10^{10} \un{s}^{-1}}{\Delta_{1}} \right)^2$ \\
Atomic motion-reduced pulse area (Raman gate): P$_{rpa}$ & \ref{sec:atomicmotion} & $ \frac{1}{ 12} \frac{\hbar^2 a^2}{mw_0^4 U_L}$ \\
Detuning Doppler shift (Raman gate): P$_{rm}$ & \ref{sec:atomicmotion} & $0.98 \left(\frac{2 \pi}{\lambda_R \Omega_R}\right)^2 \frac{\hbar}{m^{3/2} a} U_L^{1/2}$ \\
Polarization effects (Raman gate): P$_{rp}$ & \ref{sec:polarization} & $\frac{\hbar \lambda_R^2 a}{\pi w_0^4 } \frac{1}{\sqrt{2 mU_L}}$ \\
Laser intensity noise (Raman gate): P$_{rl}$  & \ref{sec:lasernoise} & $(2.6 \cdot 10^{-6} \un{m}^2 \un{s}^{-1}) \frac{1}{|\Delta_1| w_{0}^2}$
\end{tabular}
\end{ruledtabular}
  \centering 
  \caption{Error Per Gate (EPG) due to various sources. It is assumed that atoms are in the vibrational ground state (i.e., $n_x=n_y=0$). The first two sources listed produce storage errors, while the rest cause gate errors.}\label{table:epg}
\end{table*}

\begingroup
\squeezetable
\begin{table*}[ht]
\begin{ruledtabular}
\begin{tabular}{lccl}
\textbf{Source} & \textbf{Section} & \textbf{EPG} & \textbf{Numerical EPG} \\
\hline
Raman scattering (blue-detuned lattice) & \ref{sec:latticescatter} & $T_1 \frac{N}{n_A}\frac{\pi c \epsilon_0}{a \omega_L}  \sqrt{\frac{U_L(\omega_L,E_0^2)}{2 m}} \frac{\sigma(\omega_L)}{\left| \alpha(\omega_L)\right|} $ &  $2.4 \times 10^{-4}$ \\
\hline
Off-resonant transitions (microwave gate): P$_{mo}$ & \ref{sec:offresonant} & $\frac{N}{n_A} \left(\frac{\pi}{2} \frac{\delta_{T_1}}{T_1} \right)^2$ & $4.3 \times 10^{-8}$ \\
Addressing beam-induced heating (microwave gate): P$_{mh}$ & \ref{sec:addressheat} & $\frac{1}{64 \pi^4} \frac{\hbar^2 \Delta_{ac}^2 m a^6}{T_1^2 U_L^3 w_0^4}$ & $1 \times 10^{-6}\ (\ast)$ \\
Raman scattering (microwave gate): P$_{ms}$ & \ref{sec:addressscatter} & $3.4 \times 10^{-6}\ \Delta_{ac} T_1$ & $5.2 \times 10^{-5}$\\
Addressing beam position (microwave gate): P$_{mph}$ & \ref{sec:position} & $ \frac{\sqrt{2}}{\pi^3} \frac{\hbar \Delta_{ac}^2 \delta_x^2 a^5 m^{3/2}}{T_1^2 U_L^{5/2}  w_0^4}$ & $2 \times 10^{-5}\ (\ast)$ \\
\end{tabular}
\end{ruledtabular}
  \centering 
  \caption{Error Per Gate (EPG) due to various sources. In calculating a numerical value of the EPG, we used the parameters $N=10^6, n_A=100, \delta_{T_1}=10^{-10}\un{s}, \delta_x = 0.01\ \mu\mathrm{m}, \omega_L = 2 \pi c / 800\ \mathrm{nm}$, with the other parameters as defined as in Sec. \ref{sec:microwavesinglequbit} ($a = 5\un{\mu m}, U_L=200 \un{\mu K}, T_1=76\ \mu\mathrm{s},\Delta_{ac}=0.2 \un{MHz}, w_0 = 1.2 \un{\mu m}$) Where indicated above by ($\ast$), we use numerical values obtained from simulations rather than from the analytic approximations listed. We treat heating ``events'' as errors, although it is worth noting that if a re-cooling mechanism is implemented, the effect of such errors can be reduced.}\label{table:epgMicrowave}
\end{table*}
\endgroup

Having quantified the various sources of error in the previous section (summarized in Table \ref{table:epg}), we now make use of 
these results to analyze the extent to which single-qubit operations can be performed in parallel in an optical lattice.   
We shall first determine an estimate of the lower bound on the error rate due to single qubit gate errors for the case of a blue-detuned lattice with Raman transition-based single qubit gates.  We then discuss the implications of these gate errors and lattice based storage errors, together with lattice addressability issues for the scalability of quantum computation in three- and two-dimensional lattices.  Following this analysis of Raman gates, we then discuss how the corresponding arguments apply in the case of the microwave single qubit gate.

\subsection{Raman-based single qubit gate errors}
We sum the EPG for each relevant error mechanism to obtain a total error rate for the gate.  Using \req{eq:omR} and the expression for the intensity of a Gaussian beam of power $P_R$, we see that we can replace $|\Omega_R|$ by $(8.3\cdot 10^{12}  \un{W}^{-1}  \un{m}^2\un{s}^{-2})  \frac{2 P_R}{\pi w_0^2 |\Delta_1 |}$.
We also require that the Raman beam laser power $P_R$ not exceed some realistic value $P_{\mathrm{max}}$. We then numerically minimize the resulting expression for the total Raman gate error over the parameters $a$, $w_0$, $\Delta_1$, $U_L$, and $\Omega_R$ (see Raman gate terms in Table~\ref{table:epg}. Carrying out this procedure, we find that 
even if we choose an exceedingly high value of Raman laser power, $P_{\mathrm{max}} = 10 \un{W}$ and allow the other parameters to attain unrealistic values, we do not obtain a minimum EPG below $10^{-7}$ for the combined total Raman gate error mechanisms listed in Table \ref{table:epg}.  In fact, this minimum value is already unrealistically low since at these laser powers ionization of the Cs atoms would also be expected to come into play.

\begin{figure}[pt] \centering
\includegraphics[width=3.0 in,  keepaspectratio=true]{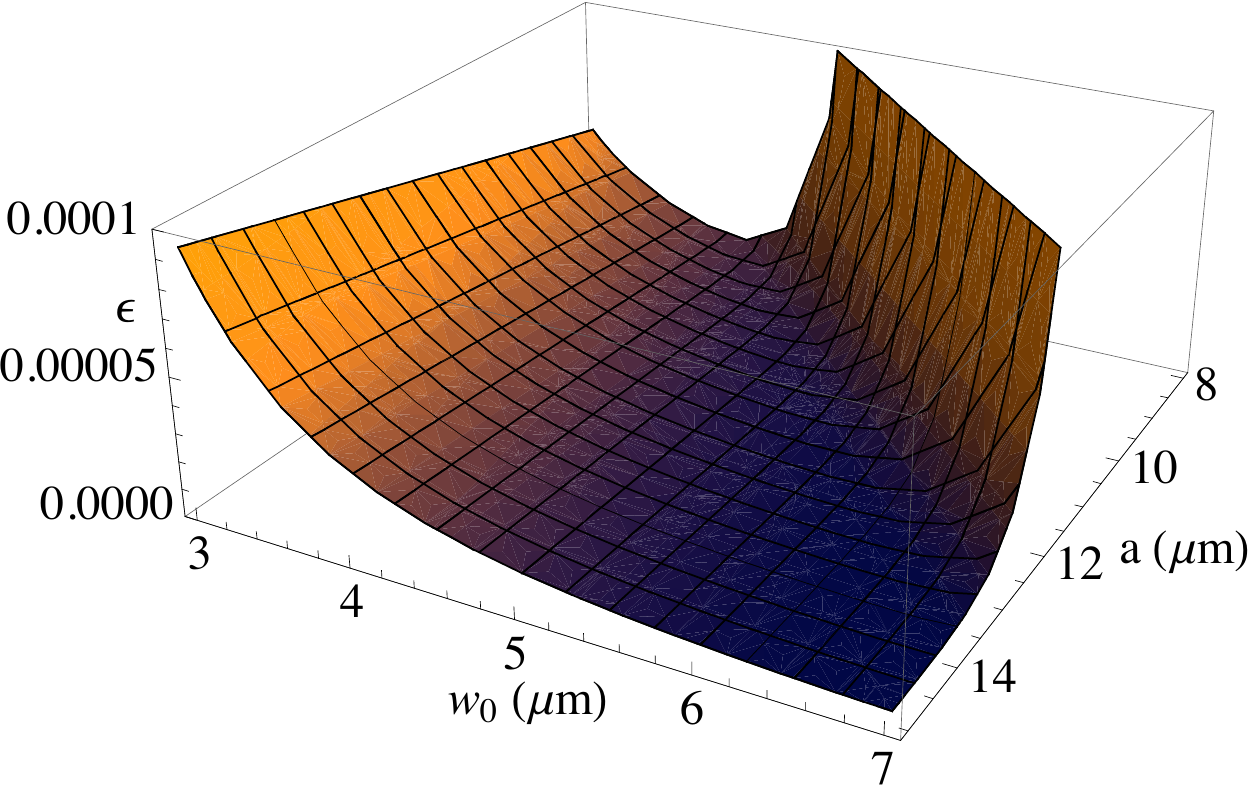}
\caption{\label{fig:optimize} (Color online) A 3D plot of gate error ($\epsilon$) versus lattice spacing ($a$) and beam waist ($w_0$) for the Raman single qubit gate, with lattice depth $U_L = 500 \un{\mu K}$ and detuning $\left| \Delta_1 \right| = 2 \pi \cdot 5 \un{THz}$.}
\end{figure}

If we now use a more realistic value of Raman laser power of $P_{\mathrm{max}} = 10 \un{mW}$, and impose realistic constraints on the other parameters, corresponding to $a < 10 \un{\mu m}$, $U_L < 500 \un{\mu K}$, $|\Delta_1| < 2 \pi \cdot 5 \un{THz}$, we
obtain a minimum EPG of $1 \times 10^{-5}$. This minimum occurs when the aforementioned parameters reach their constraint values and $w_0 =  5.0 \un{\mu m}$.  This EPG is roughly an order of magnitude better than can be achieved with our reference parameters of $a=5 \un{um}$, $U_L = 200 \un{\mu K}$ from Section~\ref{sec:intro}. A plot of the EPG as a function of $a$ and $w_0$ is shown in Fig. \ref{fig:optimize}. In this Raman-based single qubit gate scenario, polarization effects are the dominant source of gate errors, and the minimum achievable EPG is consequently most sensitive to changes in the lattice spacing, $a$. A doubling of $a_{\mathrm{max}}$ to $20 \un{\mu m}$ (and appropriate adjustment of $w_0$) results in an EPG that is approximately threefold smaller. It is also worth noting that although the EPG is sensitive to the laser power at the unrealistically high values discussed above, it is not sensitive to $P_{\mathrm{max}}$ in this regime.

This analysis suggests that one consider whether it is possible to further increase the lattice spacing $a$ beyond the maximum value $10 \un{\mu m}$ specified above.  While in principle, this would improve (reduce) the EPG, we note that the laser power at a particular frequency required to generate an optical lattice of given depth and a given number of atoms scales with the square of $a$. Thus scaling to larger lattice sizes will necessarily entail significant increases in laser power that may not be realistic.  
We note that while larger lattice spacings are beneficial for the single qubit Raman gate, they  will result in slower gate times for two-qubit gates, as a result of the weaker interaction strengths for long-range interaction-based gates such as the Rydberg gate and of longer travel distances in the case of collisional gates.

\subsection{Lattice size and scaling with Raman-based single qubit gates}
We now examine the implications of this analysis of the total Raman-based single qubit gate EPG for lattice size and scaling.  

\subsubsection{3D lattices\label{3DlatticeScale}}
With three sets of  $10\un{W}$ laser beams, we can produce a 3D lattice of $\sim 100 \times 100 \times 100$ sites, a lattice spacing of $10 \un{\mu m}$, and a lattice depth of $\sim 500 \un{\mu K}$, by tuning the laser to $\sim 851.7 \un{nm}$ (which is very close to the $D_2$ transition at $852.1 \un{nm}$). With this small a detuning, there is a substantial and undesirable $\sim 3\%$ difference in trap depth for atoms in the $\ket{0}$ state versus the $\ket{1}$ state.
This mismatch could cause entanglement of motional and internal qubit degrees of freedom and should be avoided.
Rayleigh scattering would also cause rapid heating of the trapped atoms, requiring frequent re-cooling operations. There is thus little if any room to decrease detuning further.  Consequently, larger lattices with the same trap depth that might allow a lowering of the EPG according to our analysis above would be possible only by making substantial increases in laser power. 

What are the implications of these results for the computing power of a 3D optical lattice-based scheme? Obviously, given some limit on the available lattice laser power, the lattice size and thus the number of physical qubits will be limited, with the number of qubits scaling as $\sim P_{\mathrm{max}}^{3/2}$. Another constraint is the degree of parallelizability of performing gates in an optical lattice.  As discussed earlier in Section~\ref{sec:error_mech}, fault-tolerance threshold theorems assume maximal parallelizability~\cite{steane:042322}, which implies that all or nearly all qubits need be addressable simultaneously ($n_A \simeq N$, where $n_A$ is the number of addressable qubits and $N$ is the total number of physical qubits). 
In the case of 3D addressable optical lattices, one might imagine using micro-lens arrays~\cite{birkl:microfabricated} to allow simultaneous focusing of many addressing beams. It may initially appear possible to address $\sim N^{2/3}$ atoms in a 3D lattice with this approach, using arrays of $\sim N^{1/3} \times N^{1/3}$ lenses placed  adjacent to the lattice. However, these lenses must be able to focus on sites deep inside the lattice.  Since the linear size of the lattice and thus the distance to such sites scales as $N^{1/3}$, the diameter of the lenses must also scale linearly with $N^{1/3}$. This implies that the number of addressable atoms $n_A$ in a 3D lattice is more or less constant, and does not scale with $N$. We note that 2D optical lattices (see below) do not necessarily have the same limitation, since the distance from the micro-lens array to the focal plane would be constant rather than scaling with the linear lattice dimension. 

\begin{figure}[pt] \centering
\includegraphics[width=3.0 in,  keepaspectratio=true]{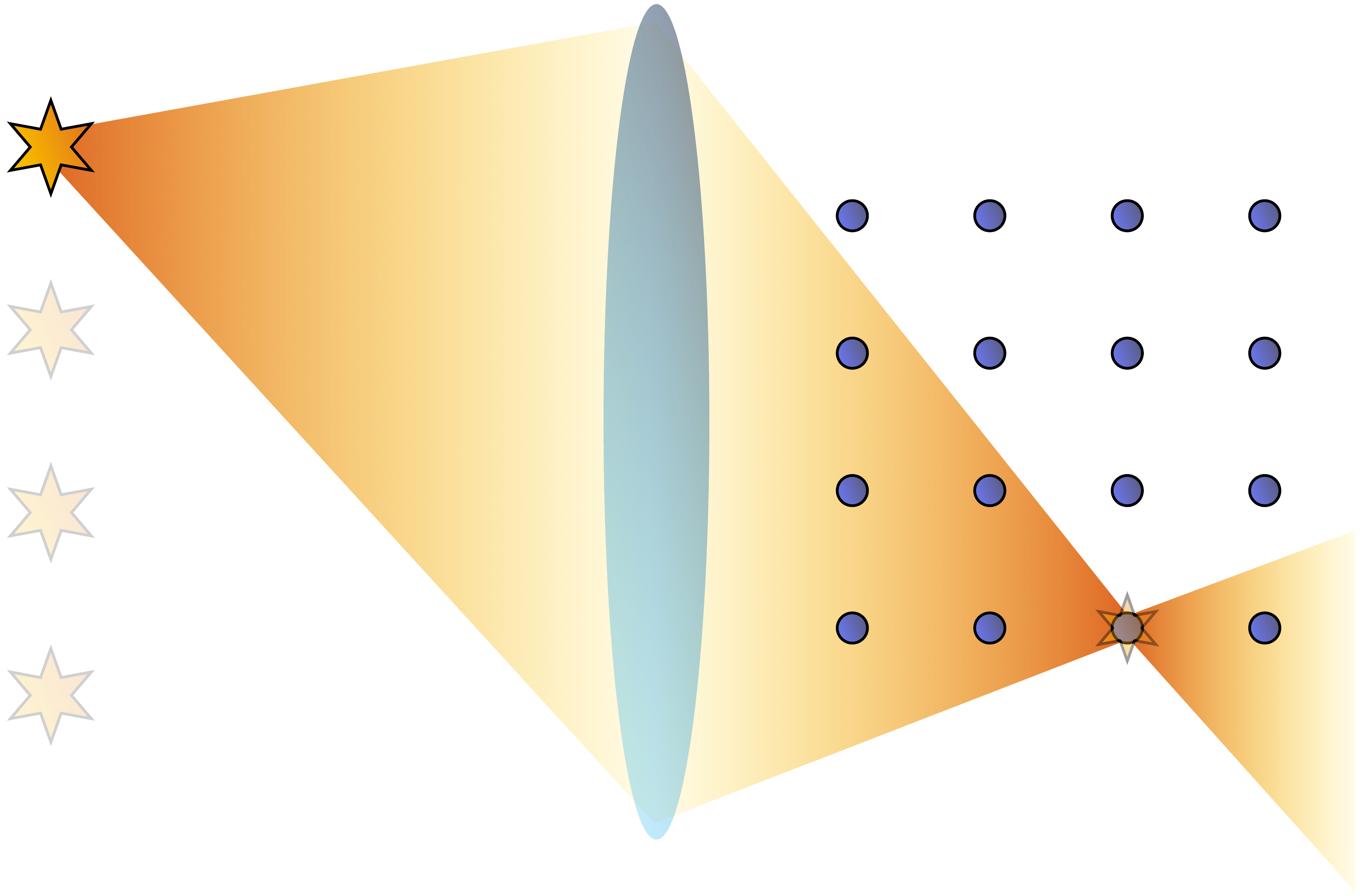}
\caption{\label{fig:biglens} (Color online) A method for addressing some or all of the atoms in a single 2D plane of a 3D optical lattice. The circles on the right represent atoms in an optical lattice (only a cross-sectional plane of atoms is shown). The stars on the left represent ``point sources'' of light, which could consist of lenses on the end of individual optical fibers, with each fiber coupled to a light source and independently controlled. The lens in the center focuses the light from the point sources onto corresponding atoms in a plane (i.e., the vertical plane perpendicular to the page surface). By adjusting either the location of the lens or the array of point sources, different planes of atoms could be targeted. }
\end{figure}

Another approach for 3D lattices would allow addressing of up to $N^{2/3}$ sites in parallel, using a single large lens and a 2D array of light sources (see Fig. \ref{fig:biglens}).
It is in principle possible to address up to all $\sim N$ atoms using a setup similar to Fig. \ref{fig:biglens} but with the ``point sources'' replaced by a spatial light modulator. However, the resolution (total number of pixels) of the spatial light modulator would have to scale with $N$, 
posing considerable challenges for implementing such an approach.

Given these constraints, how large a computation could be performed with a 3D lattice using current technology? To address this question, let us consider the aforementioned $100 \times 100 \times 100$--site lattice (generated with three $10 \un{W}$ beams). In this scenario, the per-atom Raman scattering rate for lattice light is approximately $0.4\un{s}^{-1}$. If we use $1\un{W}$ Raman beams, we can perform single qubit gates in approximately half a nanosecond on up to $10^4$ qubits in parallel. This means that the single qubit lifetime due to Raman scattering of lattice light is approximately $10^9$ time-steps, and that $10^7$ gates per qubit could be performed in that time. 

We see that in such a scenario, the storage error rate due to Raman scattering rate of lattice light are almost two orders of magnitude smaller than the gate error rate, which we calculated above to be $EPG = 1 \times 10^{-5}$. It is useful to compare these numbers with the most optimistic rigorous threshold result for local gates in a 2D architecture, which is currently $1.85 \times 10^{-5}$, with storage errors an order of magnitude smaller~\cite{svore:0604090}.  This comparison is complicated by the fact that we are considering a 3D lattice and not a 2D architecture.  The use of a 3D lattice implies a somewhat better threshold due to the reduced average distance between qubits in 3D.  Overall, this analysis suggests that Raman gates may indeed meet the fault tolerance threshold within the assumptions of our analysis (i.e., realistic laser power), if parallelizability limitations can be overcome.

\subsubsection{2D lattices}

2D lattices are subject to the same basic size constraint in terms of laser power. This reduces the number of qubits accessible, relative to a 3D lattice formed with equivalent power.  Thus, a 2D lattice with $\sim10^6$ physical qubits can be formed in a $1000 \times 1000$--site lattice with spacing $a \sim 10 \un{\mu m}$ and depth $U_L \sim 500 \un{\mu K}$.   It is straightforward to show from the analysis above that this does however now require kilowatts of power, even with the small lattice light detuning given earlier. A $100 \times 100$ lattice with the same depth and spacing could be created with more realistic $\sim10 \un{W}$ laser beams, and this would then accommodate only $\sim10^4$ physical qubits, in contrast to the $10^6$ that were possible with the 3D lattice above.

2D lattices are more attractive from the perspective of simultaneous addressability.  A micro-lens array or the method described in Fig.~\ref{fig:biglens} can in principle be used for addressing lattice sites in a 2D lattice in such a way as to allow the number of simultaneously-addressable sites $n_A$ to scale linearly with the total number of of lattice sites $N$. 
This is possible for a 2D lattice because such methods allow for parallel addressing of at most all the atoms in one plane, and a 2D lattice of course consists only of a single plane of atoms. It was recently shown that multiple standing-wave fields can also be used to address a periodic subset of the atoms~\cite{cho:020502}.

Gates performed in a 2D lattice will also not be subject to all the same error mechanisms that we would find in a 3D lattice. In particular, in a 2D lattice it may be possible to use a first-order Doppler-free configuration for the Raman beams, which would substantially reduce the detuning Doppler shift error. Nevertheless, for the Raman-based single qubit gates, polarization effects are the dominant error mechanism, and these remain more or less unchanged by the switch to a 2D lattice. Thus we still obtain an EPG value of  $1 \times 10^{-5}$ using the same parameters given above for the 3D lattice analysis.  As noted there, if no restrictions on laser power existed, we could increase the lattice spacing $a$ and the beam waist $w_0$ to reduce the EPG. In practice, laser power is likely to be the limiting factor on scaling of 2D lattices of the type considered here.

\subsection{Lattice scaling with microwave-based single qubit gates}
Our scaling analysis above has focused on the Raman single qubit gate. The microwave gate is subject to the same fundamental constraint on scaling in a 3D lattice, since it also requires a focused beam to address individual lattice sites. Microwave-based gates are typically substantially slower than Raman gates, taking on the order of tens of microseconds. This implies that lattice light scattering rates would need to be lower for a fault tolerant quantum computer employing microwave-based gates, which could be achieved with larger lattice light detuning and increased laser power. On the other hand, lattice spacings need not be quite as large for microwave-based gates, reducing the laser power required for a lattice with a given number of sites. As the former effect is larger than the latter, we expect systems of a given number of qubits employing microwave-based single qubit gates to require more power than systems of a similar number of qubits using Raman single qubit gates. 

For example, using the parameter values given in Table \ref{table:epgMicrowave} for a microwave single qubit gate-based scheme, three 75 W lasers would be required to create a $100\times100\times100$ lattice of depth $U_L = 200\ \mathrm{\mu K}$ with spacing $a=5\ \mathrm{\mu m}$ and a lattice laser wavelength of $\lambda_L = 800\ \mathrm{nm}$; despite the large detuning, atoms in such a lattice would still suffer from a large effective EPG of $2.4\times 10^{-4}$ from Raman scattering of lattice light, due to the slow gate times and limited parallelizability of microwave gates. Note that this error rate for scattering of lattice light is substantially larger than the total error rate of $\sim7\times10^{-5}$ for effects intrinsic to the microwave gate itself. By comparison, a Raman single qubit gate-based setup with the same number of lattice sites could make due with laser powers of 10 W per laser and still have an effective EPG for lattice light-induced Raman scattering that is five orders of magnitude smaller than that for the microwave gate scenario (see Sec. \ref{3DlatticeScale} for details).

While Raman single qubit gates are likely to be preferable for scaling to very large system sizes, microwave-based gates may nevertheless prove easier to implement, due to the simpler optical requirements. In particular, beam alignment is simplified in the case of microwave gates, and since they require only a single laser, they will be less sensitive to alignment errors than the Raman single qubit gates.

\section{Conclusion\label{sec:conclusion}}
We have investigated the effects of single qubit gate errors and lattice storage errors for neutral atom qubits trapped in addressable optical lattices and analyzed  the role of these errors and of other technical factors such as finite laser power in limiting the size of a quantum computation.  We find that under realistic current limitations on laser power, computation with lattices containing up to 10$^6$  qubits in three dimensions and up to $10^4$ qubits in two dimensions may be achievable.  Considerations of parallelizability and optical access are seen to impose additional limits on the scalability of such quantum computation with individually addressed gates.    These constraints are more severe for a 3D lattice than for a 2D lattice, where it is comparatively easier to develop technologies for addressing all or nearly all atoms in parallel (i.e., $n_A \simeq N$).  

Our  
quantitative scaling analysis has not included the effects of two qubit gate errors and measurement errors.   For the single qubit gates, we compared the accuracy of implementation for
two candidate gates: the stimulated Raman two-photon gate, and the microwave gate with AC Stark Shift addressing beam.

We find that Raman-based single qubit gates can be implemented for $^{133}$Cs in times on the order of a nanosecond with an error rate of $\sim1 \times 10^{-5}$ and that microwave-based single qubit gates can be implemented in times on the order of $100\ \mathrm{\mu s}$ with an error rate of $\sim7\times 10^{-5}$. (Note that neither error rate includes scattering of lattice light, which is a more serious limitation for the microwave gates due to their slower gate times.) The microwave gates are simpler to implement (e.g., require less laser alignment), but are more severely limited by constraints on site-specific parallelization due to the global nature of the microwave pulse.  Consequently, microwave gates appear to be a viable intermediate option for testing single qubit gates and realizing small scale quantum algorithms, although Raman-based gates may be preferable for eventual operation at the full capacity of addressable optical lattices.  The gate error for Raman-based single qubit operations is very close to the current best estimate for the fault tolerant threshold for computation using local gates in a 2D architecture.  The latter would be expected to be somewhat higher in a 3D architecture, suggesting that further innovations in error correction protocols may make reasonably large computations, i.e., with up to 10$^6$ qubits, achievable with 3D addressable optical lattices. 

The focus of this work was a realistic analysis of technological limitations on scalability of fault tolerant quantum computation with neutral atoms in addressable optical lattices.  Our analysis identified laser power and parallel addressability as primary factors that will eventually restrict the number of qubits for this implementation of quantum computation.   We hope that this detailed analysis will stimulate similar investigations of physical and technological limits to other proposed implementations of fault tolerant quantum computation, each of which has very different limiting physical features whose effect on scalability needs to be examined in detail.

It may also be possible to overcome the limits identified here for scalable quantum computation in 3D lattices, by making use of methods other than single-site focused beams in order to address large numbers of atoms in parallel.  Thus, entire planes of atoms can be addressed using a magnetic field gradient~\cite{schrader:150501}.  Alternatively, periodically-spaced arrays of atoms can be addressed with a second optical lattice of larger lattice constant (a ``super-lattice'')~\cite{peil:051603,cho:020502}. Both of these techniques offer high degrees of parallelization at the cost of flexibility in choosing which atoms are simultaneously addressed. Research on quantum cellular automata~\cite{raussendorf:022301} and global control schemes~\cite{kay:global} has shown that such a constraint on simultaneous addressing does not necessarily preclude efficient quantum computation, but further work is needed to understand the effect on error correction protocols and to determine fault tolerance thresholds for such schemes. 

Finally, we note that despite the limitations established here for performing large scale quantum computations, optical lattice-based schemes may also be very useful for simulating other quantum systems~\cite{jane:dynamics}. Quantum simulations have different requirements than fault-tolerant quantum computation, generally requiring less stringent accuracy of quantum operations.  Optical lattice-based quantum computers may also be useful for small-scale quantum information processing, such as a quantum repeater in a quantum key distribution network~\cite{briegel:5932}.  The long coherence time of optically-trapped neutral atoms is particularly valuable for this application.   

\begin{acknowledgments}
The authors thank Dave Weiss for critical reading of the manuscript. T.R.B. thanks Karl-Peter Marzlin, Birjoo Vaishnav, Tom Allison, and Dan Stamper-Kurn for discussions.  The effort of the authors was sponsored by the Defense Advanced Research Projects Agency (DARPA) and the Air Force Laboratory, Air Force Material Command, USAF, under Agreement No. F30602-01-2-0524. T.R.B. also acknowledges support from a Natural Sciences and Engineering Research Council of Canada (NSERC) Postgraduate Scholarship and from a US Department of Defense NDSEG Fellowship. J.V.'s current effort is sponsored by the Science Foundation Ireland through the President of Ireland Young Researcher Award. \end{acknowledgments}

\appendix*

\section{Scattering cross sections and dynamic polarizabilities in ground-state alkali atoms}

In analyzing optical lattice-based quantum computing, one frequently has to calculate optical-frequency scattering cross sections or polarizabilities for an alkali atom in a particular hyperfine sub-level of the ground state (e.g., in $^{133}$Cs, $6^2\mathrm{S}_{1/2}$, $F=3$, $m_F=1$).  Such calculations can be performed using well-known formulae; unfortunately, due to the tedious and somewhat subtle nature of these calculations, various approximations are sometimes employed as shortcuts. Such approximations are valid far from resonance, but result in significant errors when close enough to resonance to resolve the hyperfine structure, as is the case for an addressing beam at the magic wavelength, $\lambda_M$. We find it valuable to review proper methods for these calculations. 

For a quantum treatment of scattering, the relevant starting point is the Kramers-Heisenberg formula \req{eq:kh}, which can be derived using second-order time-dependent perturbation theory:

\begin{align}
d\sigma_{ab} & =  \frac{\alpha^2 \omega \omega^{\prime 3}}{c^2}  \left| \sum_i \left\{ \frac{ (\vec{x}_{bi}\cdot \vec{\epsilon}^{\prime \ast}) (\vec{x}_{ia}\cdot \vec{\epsilon}) } {\omega_{ia} - \omega} \right. \right. \nonumber \\
& \left. \left. + \frac{ (\vec{x}_{bi}\cdot \vec{\epsilon}) (\vec{x}_{ia}\cdot \vec{\epsilon}^{\prime \ast}) } {\omega_{ia} + \omega^\prime}\right\} \right|^2 d\Omega, \label{eq:kh}
\end{align}
where 
$\alpha$ is the fine structure constant, the indices $a$, $b$, and $i$ denote initial, final, and intermediate states, $\vec{x}_{ia}$ is the 
dipole matrix element and $\omega_{ia} = \omega_i - \omega_a$ the frequency for the $a \longrightarrow i$ transition, $\omega$ and $\omega^\prime$ are the frequencies of the incoming and scattered photons, and $\vec{\epsilon}$ and $\vec{\epsilon}^\prime$ are the polarization vectors of the incoming and scattered photons. In the most general case, the sum over discrete intermediate states $i$ should be extended to include an integration over continuum intermediate states, but, for our purposes we can safely neglect continuum states and other far-detuned intermediate states. Integrating over $d\Omega$ and multiplying by 2 to account for the two possible independent photon polarizations, we obtain the cross-section for a state $a$ to scatter into a state $b$ \req{eq:khsimp}:
\begin{align}
\sigma_{ab} & = \frac{8 \pi}{3} \frac{\alpha^2 \omega \omega^{\prime 3}}{c^2}  \left| \sum_i \left\{ \frac{ (\vec{x}_{ia}\cdot \vec{\epsilon}) \vec{x}_{bi} } {\omega_{ia} - \omega} + \frac{ (\vec{x}_{bi}\cdot \vec{\epsilon}) \vec{x}_{ia} } {\omega_{ia} + \omega^\prime}\right\} \right|^2. \label{eq:khsimp}
\end{align}
The Raman scattering cross section is given by $\sigma_{\text{raman},a} = \sum_{b \neq a} \sigma_{ab}$, while the Rayleigh scattering cross section is $\sigma_{\text{rayleigh},a} = \sigma_{aa}$.

Dynamic polarizability 
for an atom in state $a$ (e.g., for an AC Stark Shift) is given by \req{eq:polarizability}, which can be derived using second-order time-independent perturbation theory:
\begin{align}
\alpha_a & = \frac{e^2}{\hbar} \sum_i \left(\frac{\left| \vec{x}_{ia} \cdot \vec{\epsilon} \right|^2 }{\omega_{ia} - \omega} + \frac{\left| \vec{x}_{ai}\cdot \vec{\epsilon} \right|^2 }{\omega_{ia} + \omega} \right) \label{eq:polarizability}
\end{align}

For both \req{eq:khsimp} and \req{eq:polarizability}, the matrix elements we need to calculate are between a hyperfine sub-level $\ket{F\ m_F}$ of the $S_{1/2}$ ground state, and a hyperfine sub-level $\ket{F^\prime\ m_F^\prime}$ of either the $P_{1/2}$ or $P_{3/2}$ ($D_1$-line or $D_2$-line) excited states. Such matrix elements can be calculated using the Wigner-Eckart Theorem and measured values for the appropriate reduced matrix elements~\cite{steck:cesium, steck:rubidium}. The expression is given in the following equation \req{eq:matel}:
\begin{eqnarray}
\lefteqn{\langle F\ m_F | \vec{x} \cdot \vec{\epsilon}_q^\ast |  F^\prime\ m_F^\prime \rangle}  & & \label{eq:matel} \\
&= & \langle J || \vec{x} || J^\prime \rangle (-1)^{J+I+m_F} \sqrt{(2F+1)(2F^\prime +1)(2J+1)} \nonumber \\
& \times & \left(\begin{array}{ccc} F & 1 & F^\prime \\ m_F & q & -m_F^\prime \end{array} \right)\left\{\begin{array}{ccc} J & J^\prime & 1 \\ F^\prime & F & I \end{array} \right\}, \nonumber
\end{eqnarray}
where $\vec{\epsilon_0} = \uvec{z}$ and $\vec{\epsilon_{\pm 1}} = \mp \left(\uvec{x} \pm i \uvec{y} \right) / \sqrt{2}$ (note that $\epsilon_{\pm 1}$ corresponds to $\sigma^{\pm}$-polarized light, and also that $\epsilon_{\pm 1} = -\epsilon^\ast_{\mp 1}$). Care must be taken with polarization vectors, as in general, $\vec{x}_{ia} \cdot \vec{\epsilon} \neq \vec{x}_{ai} \cdot \vec{\epsilon}$. Thus, a particular intermediate state $i$ may contribute to the resonant term but not the anti-resonant term of \req{eq:khsimp} or \req{eq:polarizability}, or vice versa.

\end{document}